\documentclass[twocolumn,times]{aastex63}

\usepackage{graphicx}
\usepackage{makecell}
\usepackage{amsmath}
\usepackage{amsthm}
\usepackage{wrapfig}
\usepackage{booktabs}
\usepackage{lineno}
\usepackage{float} 
\usepackage{threeparttable}
\usepackage{CJKutf8}
\usepackage[squaren]{SIunits}
\usepackage{xspace}
\usepackage{comment}
\usepackage{xcolor}
\usepackage{multirow}
\usepackage{array}
\setwatermarkfontsize{1.5in}

\shortauthors{Xiao ET AL.}
\begin{document}
\begin{CJK}{UTF8}{gbsn}

\title{The peculiar precursor of a gamma-ray burst from a binary merger involving a magnetar}
\correspondingauthor{S.L Xiong, H. Gao, D. Xu and S.N Zhang}
\email{xiongsl@ihep.ac.cn, gaohe@bnu.edu.cn,\\dxu@nao.cas.cn, zhangsn@ihep.ac.cn}
\author{Shuo Xiao}
\affil{School of Physics and Electronic Science, Guizhou Normal University, Guiyang 550001, People’s Republic of China}
\affil{Key Laboratory of Particle Astrophysics, Institute of High Energy Physics, Chinese Academy of Sciences, 19B Yuquan Road, Beijing 100049, China}
\affil{University of Chinese Academy of Sciences, Chinese Academy of Sciences, Beijing 100049, China}

\author{Yan-Qiu Zhang}
\affil{Key Laboratory of Particle Astrophysics, Institute of High Energy Physics, Chinese Academy of Sciences, 19B Yuquan Road, Beijing 100049, China}
\affil{University of Chinese Academy of Sciences, Chinese Academy of Sciences, Beijing 100049, China}

\author{Zi-Pei Zhu}
\affil{Key Laboratory of Space Astronomy and Technology, National Astronomical Observatories, Chinese Academy of Sciences, Beijing 100012, China}
\affil{Department of Astronomy, School of Physics, Huazhong University of Science and Technology, Wuhan 430074, China}

\author{Shao-Lin Xiong*}
\affil{Key Laboratory of Particle Astrophysics, Institute of High Energy Physics, Chinese Academy of Sciences, 19B Yuquan Road, Beijing 100049, China}

\author{He Gao*}
\affil{Department of Astronomy, Beijing Normal University, Beijing 100875, China}

\author{Dong Xu*}
\affil{Key Laboratory of Space Astronomy and Technology, National Astronomical Observatories, Chinese Academy of Sciences, Beijing 100012, China}

\author{Shuang-Nan Zhang*}
\affil{Key Laboratory of Particle Astrophysics, Institute of High Energy Physics, Chinese Academy of Sciences, 19B Yuquan Road, Beijing 100049, China}

\author{Wen-Xi Peng}
\affil{Key Laboratory of Particle Astrophysics, Institute of High Energy Physics, Chinese Academy of Sciences, 19B Yuquan Road, Beijing 100049, China}

\author{Xiao-Bo Li}
\affil{Key Laboratory of Particle Astrophysics, Institute of High Energy Physics, Chinese Academy of Sciences, 19B Yuquan Road, Beijing 100049, China}

\author{Peng Zhang}
\affil{College of Science, China Three Gorges University, Yichang 443002, China}

\author{Fang-Jun Lu}
\affil{Key Laboratory of Particle Astrophysics, Institute of High Energy Physics, Chinese Academy of Sciences, 19B Yuquan Road, Beijing 100049, China}

\author{Lin Lin}
\affil{Department of Astronomy, Beijing Normal University, Beijing 100875, China}

\author{Liang-Duan Liu}
\affil{Department of Physics, Central China Normal University, Wuhan 430079, China}

\author{Zhen Zhang}
\affil{Key Laboratory of Particle Astrophysics, Institute of High Energy Physics, Chinese Academy of Sciences, 19B Yuquan Road, Beijing 100049, China}

\author{Ming-Yu Ge}
\affil{Key Laboratory of Particle Astrophysics, Institute of High Energy Physics, Chinese Academy of Sciences, 19B Yuquan Road, Beijing 100049, China}

\author{You-Li Tuo}
\affil{Key Laboratory of Particle Astrophysics, Institute of High Energy Physics, Chinese Academy of Sciences, 19B Yuquan Road, Beijing 100049, China}

\author{Wang-Chen Xue}
\affil{Key Laboratory of Particle Astrophysics, Institute of High Energy Physics, Chinese Academy of Sciences, 19B Yuquan Road, Beijing 100049, China}
\affil{University of Chinese Academy of Sciences, Chinese Academy of Sciences, Beijing 100049, China}

\author{Shao-Yu Fu}
\affil{Key Laboratory of Space Astronomy and Technology, National Astronomical Observatories, Chinese Academy of Sciences, Beijing 100012, China}

\author{Xing Liu}
\affil{Key Laboratory of Space Astronomy and Technology, National Astronomical Observatories, Chinese Academy of Sciences, Beijing 100012, China}
\affil{Key Laboratory of Cosmic Rays, Ministry of Education, Tibet University, Lhasa, Tibet 850000, China}

\author{Jin-Zhong Liu}
\affil{Xinjiang Astronomical Observatory, Chinese Academy of Sciences, Urumqi, Xinjiang 830011, People’s Republic of China}
\affil{School of Astronomy and Space Science, University of Chinese Academy of Sciences, Beijing 100049, People’s Republic of China}

\author{An Li}
\affil{Department of Astronomy, Beijing Normal University, Beijing 100875, China}

\author{Tian-Cong Wang}
\affil{Department of Astronomy, Beijing Normal University, Beijing 100875, China}

\author{Chao Zheng}
\affil{Key Laboratory of Particle Astrophysics, Institute of High Energy Physics, Chinese Academy of Sciences, 19B Yuquan Road, Beijing 100049, China}
\affil{University of Chinese Academy of Sciences, Chinese Academy of Sciences, Beijing 100049, China}

\author{Yue Wang}
\affil{Key Laboratory of Particle Astrophysics, Institute of High Energy Physics, Chinese Academy of Sciences, 19B Yuquan Road, Beijing 100049, China}

\author{Shuai-Qing Jiang}
\affil{Key Laboratory of Space Astronomy and Technology, National Astronomical Observatories, Chinese Academy of Sciences, Beijing 100012, China}

\author{Jin-Da Li}
\affil{Department of Astronomy, Beijing Normal University, Beijing 100875, China}

\author{Jia-Cong Liu}
\affil{Key Laboratory of Particle Astrophysics, Institute of High Energy Physics, Chinese Academy of Sciences, 19B Yuquan Road, Beijing 100049, China}
\affil{University of Chinese Academy of Sciences, Chinese Academy of Sciences, Beijing 100049, China}

\author{Zhou-Jian Cao}
\affil{Department of Astronomy, Beijing Normal University, Beijing 100875, China}

\author{Xi-hong Luo}
\affil{School of Physics and Electronic Science, Guizhou Normal University, Guiyang 550001, People’s Republic of China}

\author{Jiao-jiao Yang}
\affil{School of Physics and Electronic Science, Guizhou Normal University, Guiyang 550001, People’s Republic of China}

\author{Shu-Xu Yi}
\affil{Key Laboratory of Particle Astrophysics, Institute of High Energy Physics, Chinese Academy of Sciences, 19B Yuquan Road, Beijing 100049, China}

\author{Xi-Lu Wang}
\affil{Key Laboratory of Particle Astrophysics, Institute of High Energy Physics, Chinese Academy of Sciences, 19B Yuquan Road, Beijing 100049, China}

\author{Ce Cai}
\affil{Key Laboratory of Particle Astrophysics, Institute of High Energy Physics, Chinese Academy of Sciences, 19B Yuquan Road, Beijing 100049, China}
\affil{University of Chinese Academy of Sciences, Chinese Academy of Sciences, Beijing 100049, China}
\affil{College of Physics, Hebei Normal University, 20 South Erhuan Road, Shijiazhuang, 050024, China}

\author{Qi-Bin Yi}
\affil{Key Laboratory of Particle Astrophysics, Institute of High Energy Physics, Chinese Academy of Sciences, 19B Yuquan Road, Beijing 100049, China}
\affil{School of Physics and Optoelectronics, Xiangtan University, Yuhu District, Xiangtan, Hunan, 411105, China}

\author{Yi Zhao}
\affil{Key Laboratory of Particle Astrophysics, Institute of High Energy Physics, Chinese Academy of Sciences, 19B Yuquan Road, Beijing 100049, China}
\affil{Department of Astronomy, Beijing Normal University, Beijing 100875, China}

\author{Sheng-Lun Xie}
\affil{Key Laboratory of Particle Astrophysics, Institute of High Energy Physics, Chinese Academy of Sciences, 19B Yuquan Road, Beijing 100049, China}
\affil{Department of Physics, Central China Normal University, Wuhan 430079, China}

\author{Cheng-Kui Li}
\affil{Key Laboratory of Particle Astrophysics, Institute of High Energy Physics, Chinese Academy of Sciences, 19B Yuquan Road, Beijing 100049, China}

\author{Qi Luo}
\affil{Key Laboratory of Particle Astrophysics, Institute of High Energy Physics, Chinese Academy of Sciences, 19B Yuquan Road, Beijing 100049, China}
\affil{University of Chinese Academy of Sciences, Chinese Academy of Sciences, Beijing 100049, China}

\author{Li-Ming Song}
\affil{Key Laboratory of Particle Astrophysics, Institute of High Energy Physics, Chinese Academy of Sciences, 19B Yuquan Road, Beijing 100049, China}

\author{Shu Zhang}
\affil{Key Laboratory of Particle Astrophysics, Institute of High Energy Physics, Chinese Academy of Sciences, 19B Yuquan Road, Beijing 100049, China}

\author{Jin-Lu Qu}
\affil{Key Laboratory of Particle Astrophysics, Institute of High Energy Physics, Chinese Academy of Sciences, 19B Yuquan Road, Beijing 100049, China}

\author{Cong-Zhan Liu}
\affil{Key Laboratory of Particle Astrophysics, Institute of High Energy Physics, Chinese Academy of Sciences, 19B Yuquan Road, Beijing 100049, China}

\author{Xu-Fang Li}
\affil{Key Laboratory of Particle Astrophysics, Institute of High Energy Physics, Chinese Academy of Sciences, 19B Yuquan Road, Beijing 100049, China}

\author{Yu-Peng Xu}
\affil{Key Laboratory of Particle Astrophysics, Institute of High Energy Physics, Chinese Academy of Sciences, 19B Yuquan Road, Beijing 100049, China}

\author{Ti-Pei Li}
\affil{Key Laboratory of Particle Astrophysics, Institute of High Energy Physics, Chinese Academy of Sciences, 19B Yuquan Road, Beijing 100049, China}
\affil{University of Chinese Academy of Sciences, Chinese Academy of Sciences, Beijing 100049, China}
\affil{Department of Astronomy, Tsinghua University, Beijing 100084, China}

\begin{abstract}
The milestone discovery of GW 170817-GRB 170817A-AT 2017gfo has shown that gravitational wave (GW) could be produced during the merger of neutron star-neutron star/black hole and that in electromagnetic (EM) wave a gamma-ray burst (GRB) and a kilonova (KN) are generated in sequence after the merger. 
Observationally, however, EM property before the merger phase is still unclear. Here we report a peculiar precursor in a KN-associated long-duration GRB 211211A, providing evidence of the EM before the merger. 
This precursor lasts $\sim$ 0.2 s, and the waiting time between the precursor and the main burst is $\sim$ 1 s, comparable to that between GW 170817 and GRB 170817A. 
The spectrum of the precursor could be well fit with a non-thermal cutoff power-law model instead of a blackbody. 
Especially, a $\sim$22 Hz Quasi-Periodic Oscillation candidate ($\sim 3\sigma$) is detected in the precursor. 
These temporal and spectral properties indicate that this precursor is probably produced by a catastrophic flare accompanying with magnetoelastic or crustal oscillations of a magnetar in binary compact merger. 
The strong magnetic field of the magnetar can also account for the prolonged duration of GRB 211211A. However, it poses a challenge to reconcile the rather short lifetime of a magnetar with the rather long spiraling time of a binary neutron star system only by the GW radiation before merger.

\end{abstract}


\section{Introduction}
As the most violent explosions in the universe \citep{zhang2018physics}, GRBs are traditionally classified into long GRBs (LGRBs) and short GRBs (SGRBs) based on their durations separated at $\sim$ 2 s \citep{kouveliotou1993identification}. Some characteristics and empirical relationships, such as hardness ratio, spectral lag, minimum variability timescale and Amati relation \citep{amati2002intrinsic,zhang2009discerning} are also employed to determine whether a GRB originates from massive star core collapse, or from binary neutron star mergers \citep{abbott2017gravitational,woosley2006supernova} where a kilonova may be observed owing to r-process nucleosynthesis \citep{li1998transient,wu2017imprints}. 

About 10\% GRBs (mostly LGRBs) detected by \textit{Swift}/BAT have precursors, and no significant difference was found between the precursor and the main emission, implying that the precursor share the same physical origin with the main burst and directly originate from central engine activities \citep{hu2014internal,li2022temporal}. On the other hand, SGRBs with precursors are very rare. Only about 2.7\% SGRBs detected by \textit{Swift}/BAT and \textit{Fermi}/GBM have precursors. As of April 2020, \textit{Fermi}/GBM only detected about 16 precursors in SGRBs \citep{wang2020stringent}. 

The thermal-spectrum precursor of SGRB is usually attributed to shock breakout or photospheric radiation of a fireball launched after the merger, whereas the non-thermal precursor is mostly interpreted with magnetospheric interaction between two neutron stars \citep{hansen2001radio,troja2010precursors,palenzuela2013electromagnetic,wang2018pre} or the crust resonant shattering of neutron star \citep{tsang2012resonant} prior to the merger. As shown in the spectral analysis of SGRB precursors observed by \textit{Fermi}/GBM \citep{wang2020stringent}, there is no `decisive' evidence of a non-thermal spectrum ($\Delta BIC<6$) for any precursor yet. In another study with \textit{Swift} and \textit{Fermi}/GBM data \citep{zhong2019precursors}, some candidate precursors prefer non-thermal CPL model with $\Delta BIC>10$, but none of these events satisfies the precursor criteria strictly.
Indeed, it is very difficult to constrain the energy spectrum of SGRB precursors because they are usually very weak with limited statistics.

In this work, observations of GRB 211211A are briefly described in Section 2, and the identification, temporal and spectral analyses of the precursor are presented in Section 3. Finally, discussion and conclusion is given in Section 4 and 5, respectively.

\section{Observations}
GRB 211211A was detected at 2021-12-11 13:09:59.651 (UTC, denoted as $T_0$) by \textit{Fermi}/GBM \citep{mangan2021grb}, \textit{Swift}/BAT \citep{d2021grb} and \textit{Insight}-HXMT/HE \citep{zhang2021grb}. The full burst lightcurve can be divided into three emission episodes: a precursor (PRE) with duration of $\sim 0.2$ s in Fast Rising Exponential Decay (FRED) shape (as shown in \ref{lc_he_gbm_bat}), a $\sim 10$ s spiky hard main emission (ME), and a soft long extended emission (EE) with $> $50 s \citep{mangan2021grb}. This burst has a duration of $T_{90}\sim 34.3$ s in 10-1000 keV by \textit{Fermi}/GBM \citep{mangan2021grb}.
 
With detailed analysis of the prompt emission, we find that the spectral and temporal properties, which include small minimum variability timescale (8.5$\pm0.8$ ms) calculated by Bayesian Blocks (BB) algorithm \citep{scargle2013studies,Xiao2023}, small spectral lag ($3.6\pm 1.6$ ms between 200-250 keV and 100-150 keV) calculated by Li-CCF method \citep{li2004timescale, xiao2022robust}, as well as the positions in $T_{90} - E_{\rm peak}$ and Amati correlation diagrams \citep{Mei2022,yang2022long,troja2022nearby,Gompertz2023}, confirm that GRB 211211A originates from a compact binary merger.
 
On the other hand, no accompanied supernovae but an AT 2017gfo-like kilonova was reported for this burst \citep{rastinejad2022kilonova}, which was also confirmed by Nanshan/NEXT 0.6-m telescope. Optical follow-ups at the Nanshan/NEXT 0.6-m telescope localised the burst in a bright ($r = 19.5$) galaxy SDSS J140910.47+275320.8, for which a spectroscopic redshift of $z = 0.076$ was reported by the Nordic Optical Telescope \citep{malesani2021grb}. The offset between the burst and the nucleus of the galaxy of $5.11\ \pm \ 0.23 $ arcsec, corresponding to $ 7.61\ \pm \ 0.34$ kpc in projection, indicates a very small probability of chance coincidence of $1.1 \%$ \citep{bloom2002observed}. Considering the measured distance, the isotropic equivalent energies $E_{\rm iso}$ for PRE, ME and EE of GRB 211211A are $6.97\times10^{48} $ erg, $9.02\times10^{51}$ erg and $2.60\times10^{51}$ erg, respectively.

\renewcommand{\thefigure}{Figure. \arabic{figure}}
\renewcommand{\figurename}{}
\setcounter{figure}{0}
\begin{figure*}
	\centering
	\begin{minipage}{0.49\linewidth}
		\centering
		\includegraphics[width=1\textwidth]{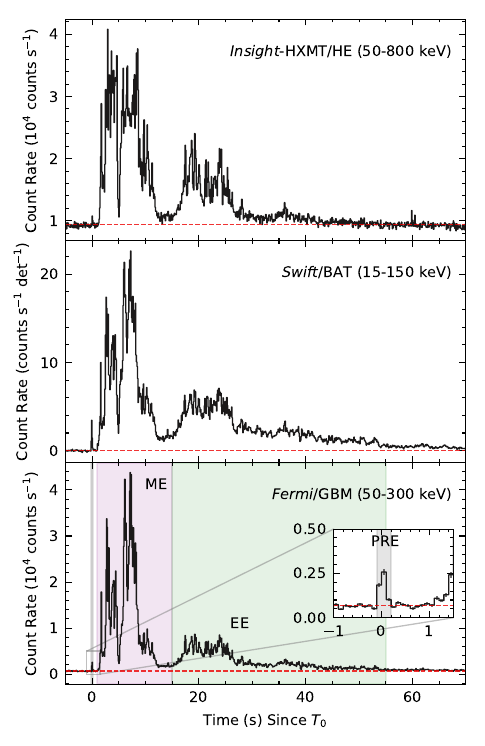}
	\end{minipage}
	\begin{minipage}{0.49\linewidth}
		\centering
		\includegraphics[width=1\textwidth]{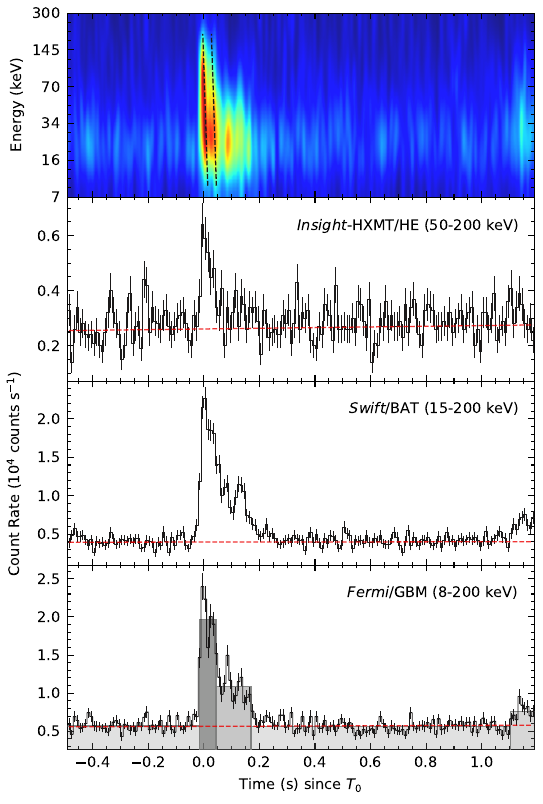}
	\end{minipage}

\caption {\textbf{Left panels:} Lightcurves of GRB 211211A. The top panel is the summed {\it Insight}-HXMT/HE lightcurve in 50-800 keV energy range (the incident energy greater than $\sim$100 keV). The second panel is the summed {\it Swift}/BAT mask-weighted and background-subtracted lightcurve in 15-150 keV energy range. The third panel is the summed {\it Fermi}/GBM lightcurve in 50-300 energy range. \textbf{Right panels:} Lightcurves of the precursor observed by \textit{Insight}-HXMT/HE, \textit{Swift}/BAT and \textit{Fermi}/GBM, the time bin is 10 ms. The light travel time difference has been corrected. The $T_{\rm pre}$ and $T_{\rm wt}$ at 8-200 keV band are about 0.19 s and 0.93 s, respectively. The shaded bars in the bottom panel show the lightcurve structure obtained by BB algorithm. The black dashed lines in the top panel represent logarithmic spectral lag behavior on time-energy domain, $t(E)=0.006\ln(E)+C$.} 
\label{lc_he_gbm_bat}
\end{figure*}

\begin{center}
\begin{figure*}
    \centering
    \begin{minipage}{0.95\linewidth}
        \setlength\tabcolsep{0.1pt}
        \begin{tabular}{c c c}
            \includegraphics[width=0.33\textwidth]{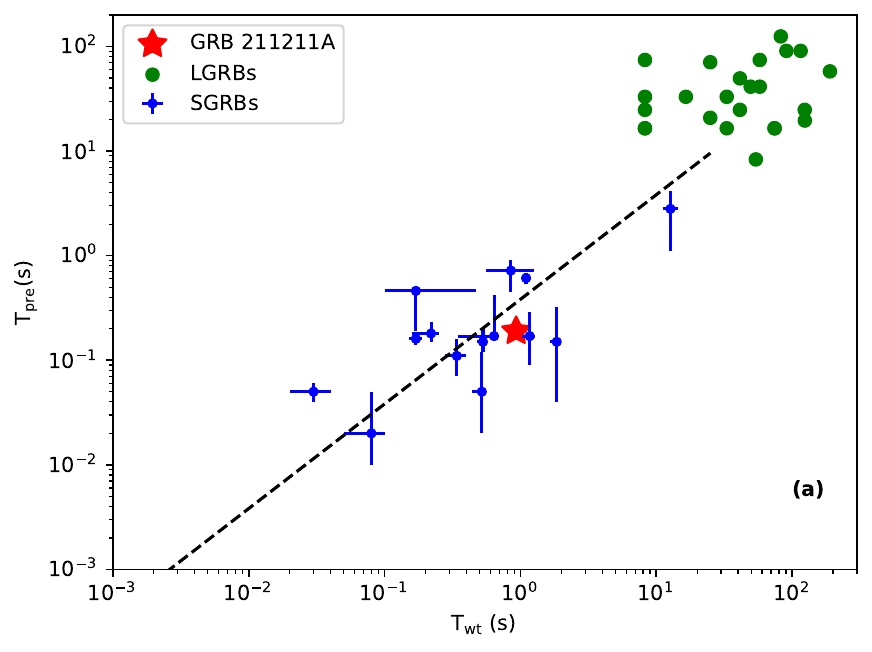}
             &
            \includegraphics[width=0.33\textwidth]{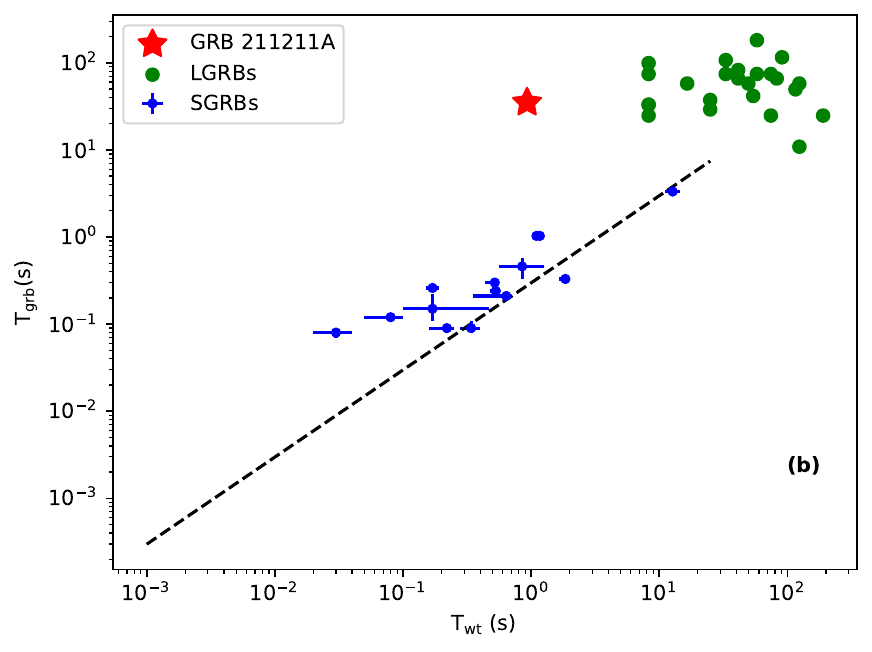}
             &
            \includegraphics[width=0.33\textwidth]{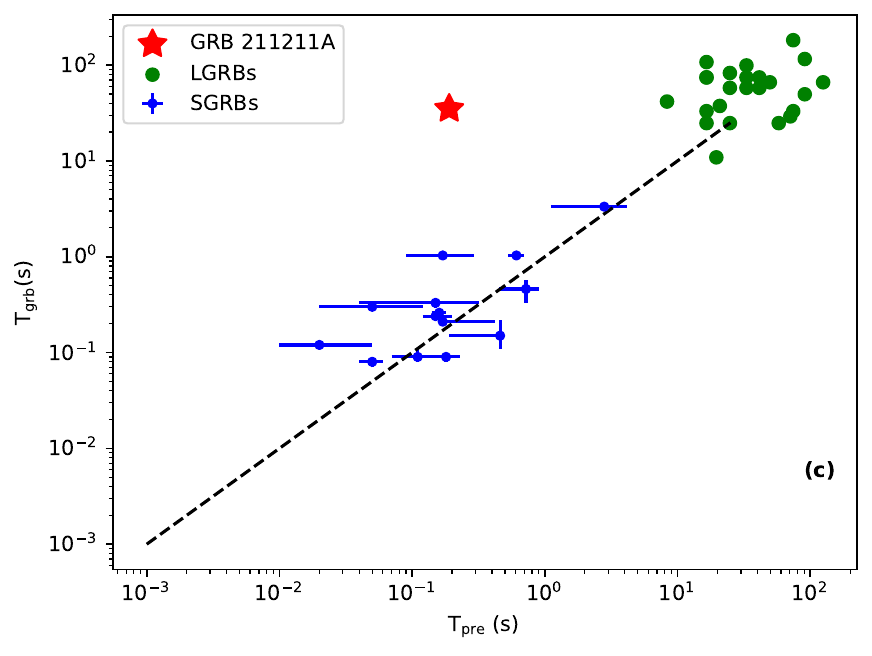}
        \end{tabular}
    \end{minipage}
\caption {\textbf{(a)} The waiting time $T_{\rm wt}$ v.s. the full duration of the precursor emission $T_{\rm pre}$. \textbf{(b)} $T_{\rm wt}$ v.s. the full duration of the GRB $T_{\rm grb}.$ \textbf{(c)} $T_{\rm pre}$ v.s. $T_{\rm grb}$. The GRB sample is obtained from the literature \citep{hu2014internal,wang2020stringent}.} 
\label{Merger_zj}
\end{figure*}
\end{center}

\section{The Peculiar Precursor}
\subsection{Identification of precursor}
A precursor should satisfy the following requirements \citep{wang2020stringent}: (1) It should be the first pulse in the light curve; (2) The peak count rate is lower than that of the main pulse; (3) The count rate during the waiting time period, i.e., the time interval between the precursor and the main pulse, is consistent with the background level; (4) It is significant enough to be identified as a burst signal. 

Bayesian Block (BB) algorithm \citep{scargle2013studies} is widely used to identify the precursor \citep{hu2014internal, wang2020stringent} of GRBs. Here we employ the same technique to confirm the existence of the precursor of GRB 211211A. Table \ref{table_bkg_test} lists results of the precursor and waiting time obtained from BB algorithm.

For the third criterion, two statistical schemes are adopted to test whether the lightcurve during the waiting time period is consistent with background. We first calculate the significance of the `possible weak signal' within $T_{\rm wt}$ by Li-Ma formula \citep{li1983analysis}. After choosing the quiescent period determined by the Bayesian Block algorithm as $t_{\rm on}$ and the confirmed background region from $T_0-1.1$ s to $T_0-0.1$ s as $t_{\rm off}$, the Li-Ma formula gives a significance of 0.5 $\rm \sigma$ for GBM and 1.7 $\rm \sigma$ for BAT, which are consistent with background fluctuation. 

To further investigate the lightcurve behavior during the waiting time period, we fit the $T_0-1.1$ s to $T_0-0.1$ s background lightcurves of three instruments with a first-order polynomial, and interpolate over $T_0-0.5$ s to $T_0+1.25$ s, then the estimate of background rate is obtained. The significance of a potential signal relative to background noise level can be calculated through weighting the difference between the lightcurve and estimated background, this quantity is commonly referred to as $\chi$. We apply the $\chi^2$ and Anderson–Darling test to examine the overall fluctuation and normality of $\chi$ within this period, respectively. As shown in \ref{table_bkg_test}, the large p-value of $\chi^2$ test implies the overall fluctuation of $\chi$ is consistent with background, and the high significance of Anderson–Darling test shows the well normality of $\chi$, i.e., the lightcurves behave like background. All the statistical evidence indicates that there exists no significant signal in the quiescent period, thus the signal near $T_0$ should be identified as the precursor of the GRB. The precursor is well separated from the main burst by quiescence period ($T_{\rm wt}$) of about 1 s, during which the lightcurve decreases to the background. We note that this is the only quiescence time interval found throughout the entire burst.

Interestingly, we find that none of GRBs with confirmed or candidate kilonova reported earlier \citep{jin2020kilonova}, including GRB 050709, GRB 060614, GRB 061201, GRB 070809, GRB 130603B, GRB 150101B, GRB 160621B, GRB 170817A, has a precursor \citep{wang2020stringent, zhong2019precursors}. 
Recently, GRB 200522A has been reported to have a precursor but the kilonova is not confirmed \citep{o2021tale}. GRB 230307A has a confirmed kilonova \citep{Gillanders2023element,Levan2023,yang2024lanthanide} but the suggested precursor candidate \citep{Dichiara2023} does not strictly meet the criteria of quiescence period between the precursor and main burst. Therefore, the precursor of GRB 211211A seems to be the only one in the GRBs with confirmed kilonova so far.

\subsection{Temporal and spectrum analysis}
The overall shape of the precursor lightcurve can be fitted well with the FRED model, with the rising time of 8.0$^{+2.4}_{-2.2}$ ms, 10.0$^{+2.1}_{-2.2}$ ms and 10.4$^{+12.2}_{-6.4}$ ms and the decay time of 84.0$^{+10.8}_{-10.7}$ ms, 99.2$^{+8.0}_{-8.4}$ ms and 43.1$^{+12.0}_{-12.3}$ ms for \textit{Fermi}/GBM, \textit{Swift}/BAT and HXMT/HE data respectively. With the BB algorithm, we calculate the duration of the precursor as $T_{\rm pre}\sim 0.19 $ s ($T_0$-0.017 to $T_0$+0.17 s) for GBM (8-200 keV), $\sim$ 0.2 s for BAT (15-200 keV), $\sim$ 0.1 s for HXMT/HE (50-200 keV), and the waiting time between the precursor and the main burst as $T_{\rm wt}\sim 0.93$ s ($T_0$+0.17 to $T_0$+1.10 s) for GBM, $\sim$ 0.88 s (BAT) and $\sim$ 1.28 s (HXMT/HE). Interestingly, this precursor follows well the relation between the waiting time and precursor duration of SGRBs (see \ref{Merger_zj}), lending additional support to the merger origin despite of the prolonged main burst. However, this precursor is outlier in $T_{\rm grb}$-$T_{\rm wt}$ and $T_{\rm grb}$-$T_{\rm pre}$ relationships (see \ref{Merger_zj}) owing to the prolonged main burst.

\begin{figure*}
\centering
\begin{minipage}[t]{0.46\textwidth}
\centering
\includegraphics[width=\columnwidth]{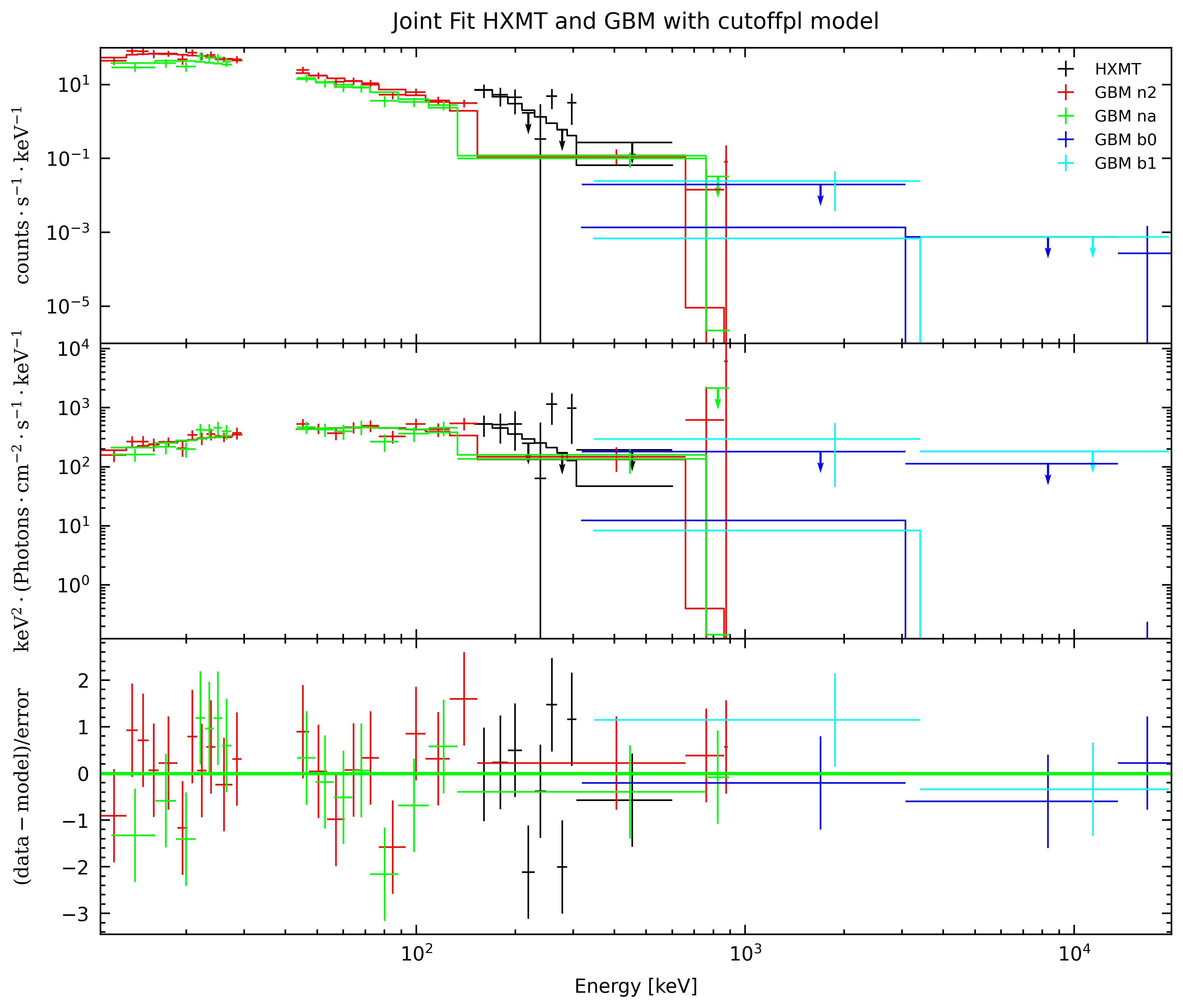}
\end{minipage}
\begin{minipage}[t]{0.46\textwidth}
\centering
\includegraphics[width=\columnwidth]{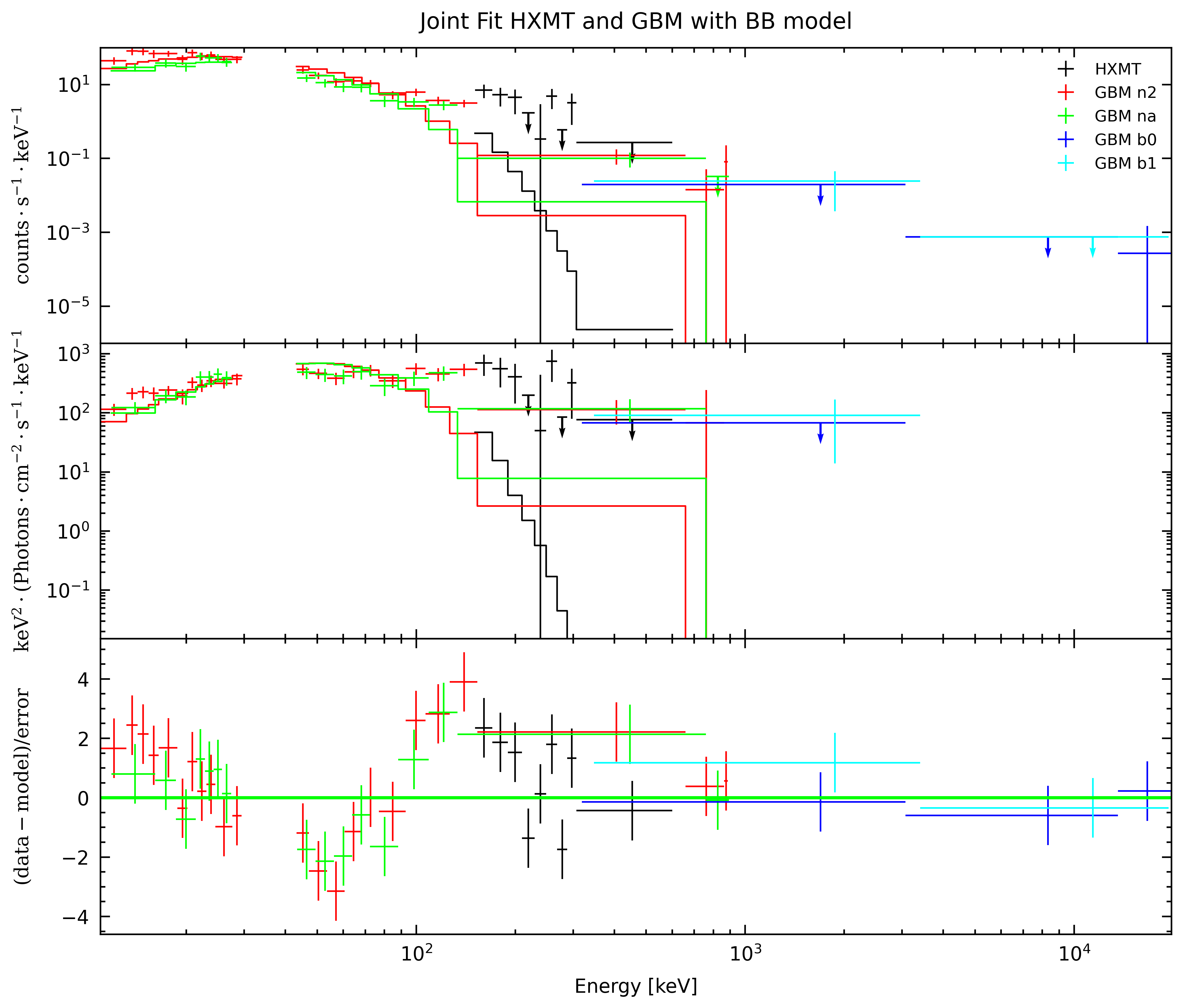}
\end{minipage}
\caption{The spectral fitting with the CPL model (left panel) and blackbody model (right panel) of the precursor with {\it Fermi}/GBM and {\it Insight}-HXMT/HE for GRB 211211A.}\label{fits_comp}
\end{figure*}

Spectrum fitting for the precursor has been performed with \textit{Fermi}/GBM and \textit{Insight}-HXMT data.
The Bayesian Information Criterion (BIC) \citep{schwarz1978estimating} is used to measure the goodness of fit and determine the best model. Comptonization (CPL), Blackbody (BB), CPL + BB models are used (see Table \ref{fits_models}). Since the CPL-only model compared to CPL + BB model yields $\Delta BIC>8$, and CPL model compared to the BB model (\ref{fits_comp}) yields $\Delta BIC>155$, we find that the best model of the precursors is the non-thermal spectrum (CPL), which displays `decisive' evidence \citep{liddle2007information} against the models with higher criterion values. 

The energy spectra of all SGRBs precursors observed by the GBM have been stringently searched and studied in literature \citep{wang2020stringent}, but none of them has `decisive' evidence of a non-thermal spectrum ($\Delta BIC<6$). On the other hand, the precursors of SGRBs observed by Swift and GBM are investigated in literature \citep{zhong2019precursors}, and although several of them prefer CPL with $\Delta BIC>10$, none of these strictly satisfies the precursor criteria, i.e., the count rate during the waiting time period is consistent with the background level (e.g. GRB 160726A) or a significance high enough to be identified as signal (e.g. GRB 180402A and bn 160818198), or it should be a LGRB (e.g. GRB 140209A). Therefore, the precursor of GRB 211211A is a rare burst confirmed with a non-thermal spectrum in the SGRBs precursors (merger origin) thanks to its high brightness.

For this precursor, the average flux is $\rm 1.76_{-1.04}^{+0.08} \times 10^{-6}$ $\rm erg/cm^2/s$, the time-averaged isotropic luminosity is $\rm 2.53_{-0.47}^{+0.06} \times 10^{49}$ $\rm erg/s$, the 64-ms peak isotropic luminosity is $\rm 7.66_{-1.19}^{+0.36} \times 10^{49}$ $\rm erg/s$ and the total isotropic energy is $\rm 6.97_{-1.29}^{+0.15} \times 10^{48}$ erg, considering the distance of 346.1 Mpc which is derived from the redshift measurement.

\begin{figure}
\includegraphics[width=\columnwidth]{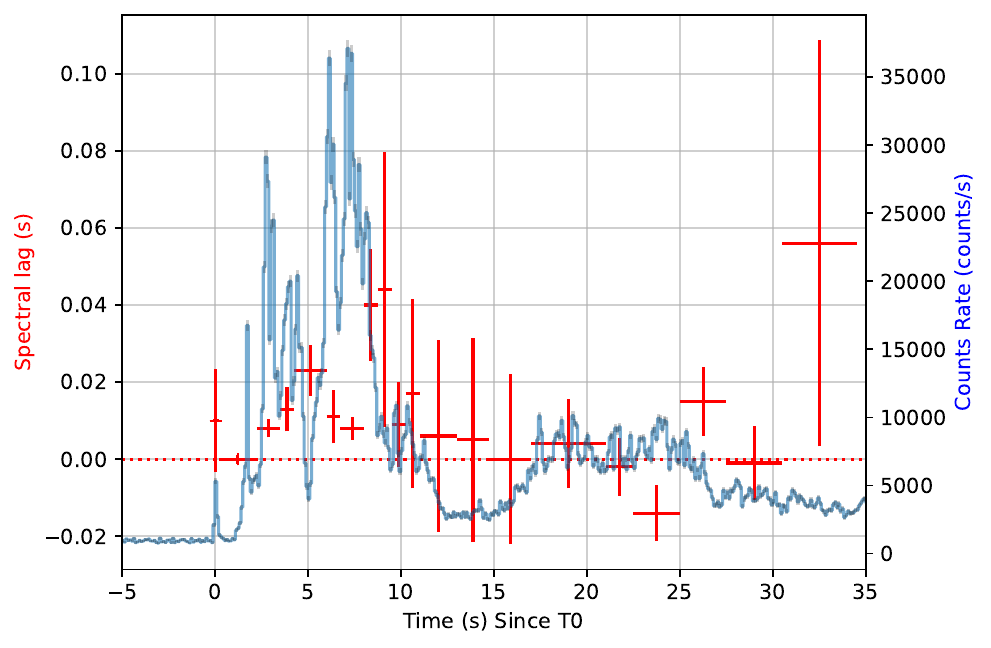}
\caption {Spectral lag (100-300 keV compared to 8-50 keV) evolution in GRB 211211A. The detector n2 with the optimal incidence angle is used. } \label{t_lag}
\end{figure}

\begin{figure}
\includegraphics[width=\columnwidth]{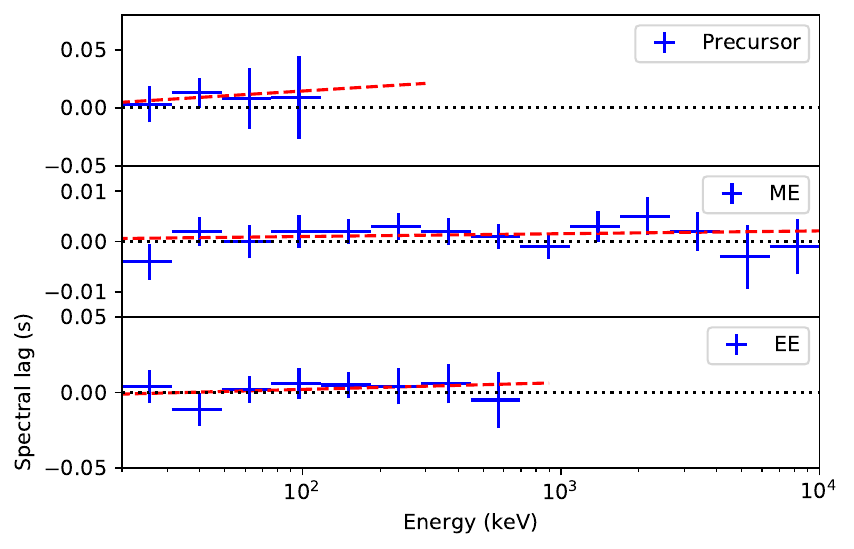}
\caption {Energy dependent spectral lag between
the lowest energy band (8-12 keV) and any higher energy band. The detectors n2 and b0 with the optimal incidence angles are used. From top to bottom panels are the spectral lags of Precursor, ME and EE, which can be fitted with $t(E)\approx0.006 (\pm 0.002)\ln(E)-0.013$ s, $t(E)\approx0.0002 (\pm 0.0003)\ln(E)-0.0001$ s and $t(E)\approx0.002 (\pm 0.002) \ln(E)-0.007$ s, respectively.} \label{lag_e}
\end{figure}

In addition, we find that spectral analysis jointly with \textit{Fermi}/GBM and HXMT/HE data show significant spectral evolution in PRE, ME and EE episodes (Table \ref{fits_table}). As shown in \ref{fits_table}, the spectrum of the precursor is significantly softer than ME.
The spectral lags (100-300 keV compared to 8-50 keV) in different time ranges observed by \textit{Fermi}/GBM are shown in \ref{t_lag}. The spectral lag of the precursor is slightly larger than that of main pulses during the ME, which is confirmed by an independent study \citep{troja2022nearby}. We further investigate the relationship between spectral lags and energy bands and obtain similar result, as shown in \ref{lag_e}. 
Besides, we find the minimum variability timescale of the precursor is $15\pm2$ ms, which is slightly larger compared to that of the main emission ($8.5\pm0.8$ ms). These temporal and spectral characteristics also indicate that the origin of this precursor should be different from the main burst.

\begin{table*}
\caption{Identification of the precursor of GRB 211211A.}
\centering
\resizebox{\textwidth}{!}{\begin{tabular}{ccccccccc}

\hline \hline
Instrument & $T_{\rm pre}$ (s) & $T_{\rm wt}$ (s) & ${b_0}^*$ (counts/s) & ${b_1}^*$ (counts/s$^2$) & $\chi^2$/d.o.f & p-value & $A^2/n$ & significance \\
\hline

\textit{Insight}-HXMT/HE & 0.08 (0.00:0.08) & 1.28 (0.08:1.35) & 2610.40$\pm$135.23 & 61.45$\pm$101.37 & 74.69/86 & 0.80 & 0.30/88 & $\textgreater$ 15 $\sigma$ \\

\textit{Swift}/BAT & 0.22 (0.00:0.22) & 0.88 (0.22:1.10) & 3985.38$\pm$165.82 & 17.21$\pm$135.69 & 73.48/86 & 0.83 & 0.27/88 & $\textgreater$ 15 $\sigma$ \\

\textit{Fermi}/GBM & 0.19 (-0.01:0.17) & 0.93 (0.17:1.10) & 5651.43$\pm$174.77 & 62.74$\pm$136.93 & 90.18/86 & 0.36 & 0.31/88 & $\textgreater$ 15 $\sigma$ \\

\hline \hline
\label{table_bkg_test}
\end{tabular}}

\begin{tablenotes}[flushleft]  
\centering
\item{$^*$} $b_0$ is the background rate at $T_0$, and $b_1$ is the background change rate.
\end{tablenotes} 
\end{table*}

\begin{figure*}
	\centering
	\begin{minipage}{0.46\linewidth}
		\centering
        \includegraphics[width=8.5cm,height=6cm]{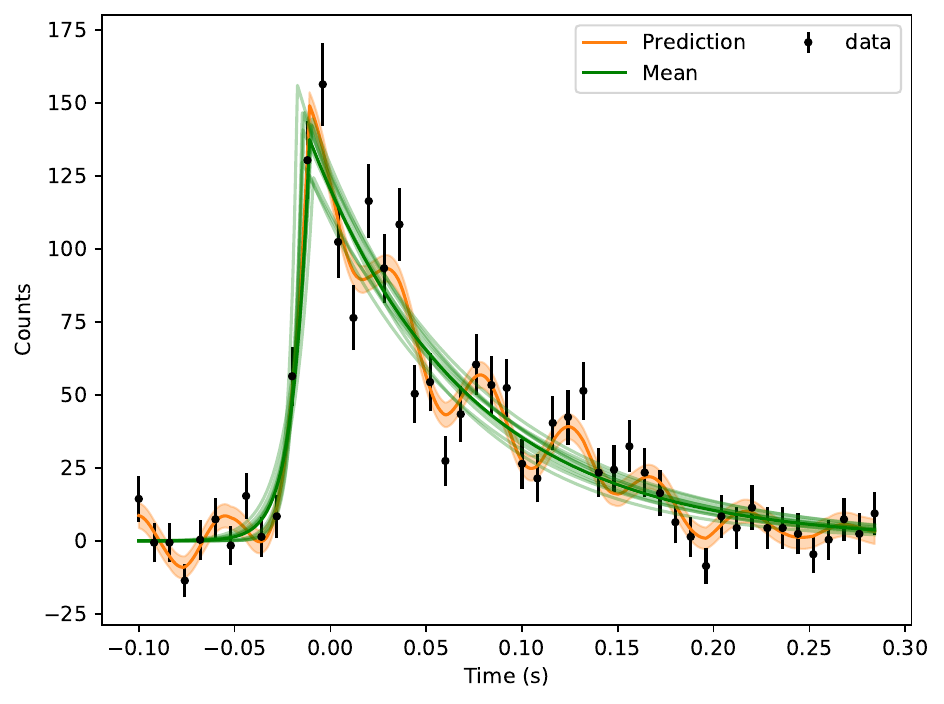}
	\end{minipage}
	\begin{minipage}{0.46\linewidth}
		\centering
        \includegraphics[width=8.5cm,height=6cm]{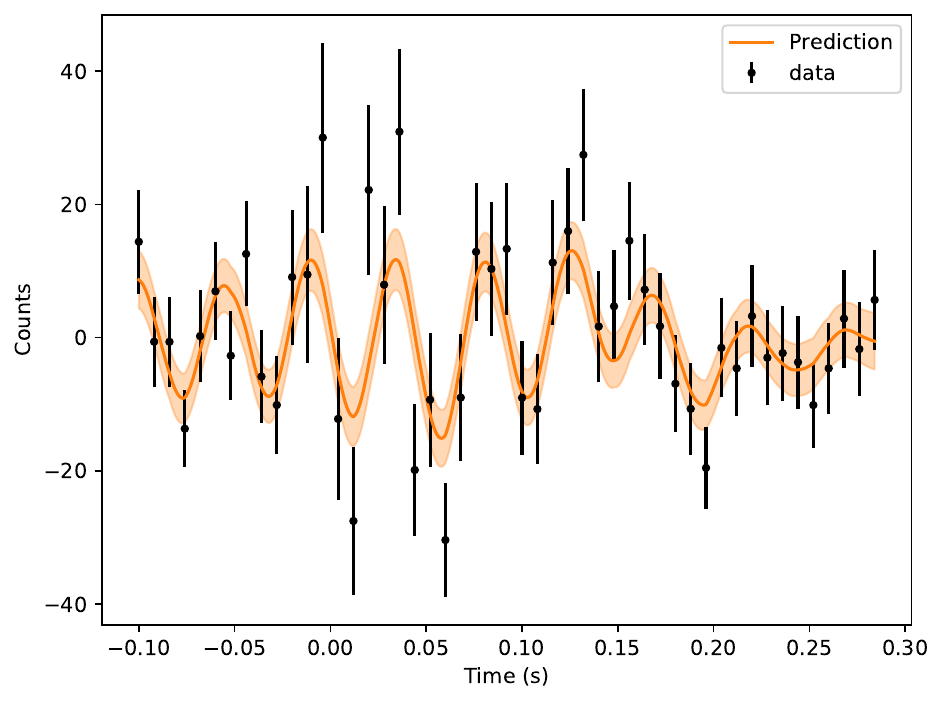}
	\end{minipage}

	\begin{minipage}{0.46\linewidth}
		\centering
  	\includegraphics[width=8.5cm,height=6cm]{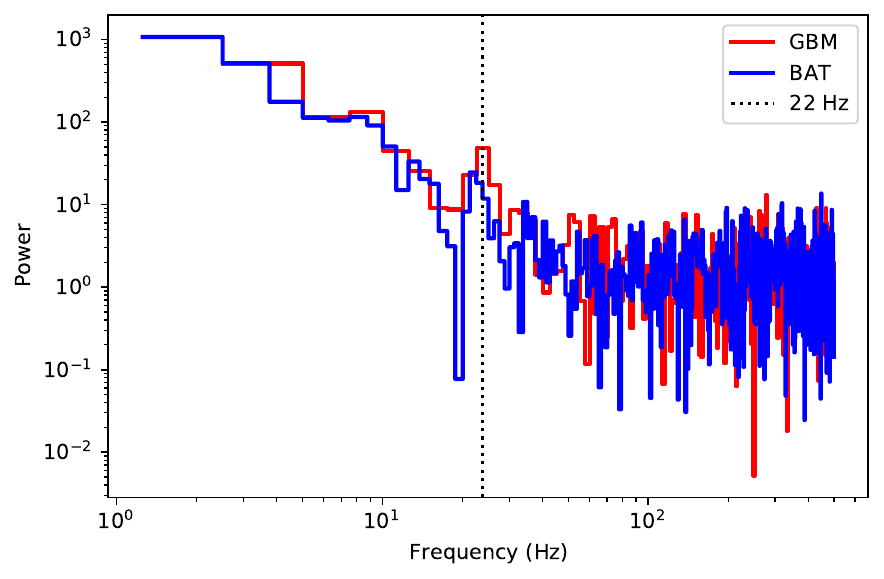}
	\end{minipage}
	\begin{minipage}{0.46\linewidth}
        \vskip -1.0cm
		\centering
  	\includegraphics[width=9cm,height=6.41cm]{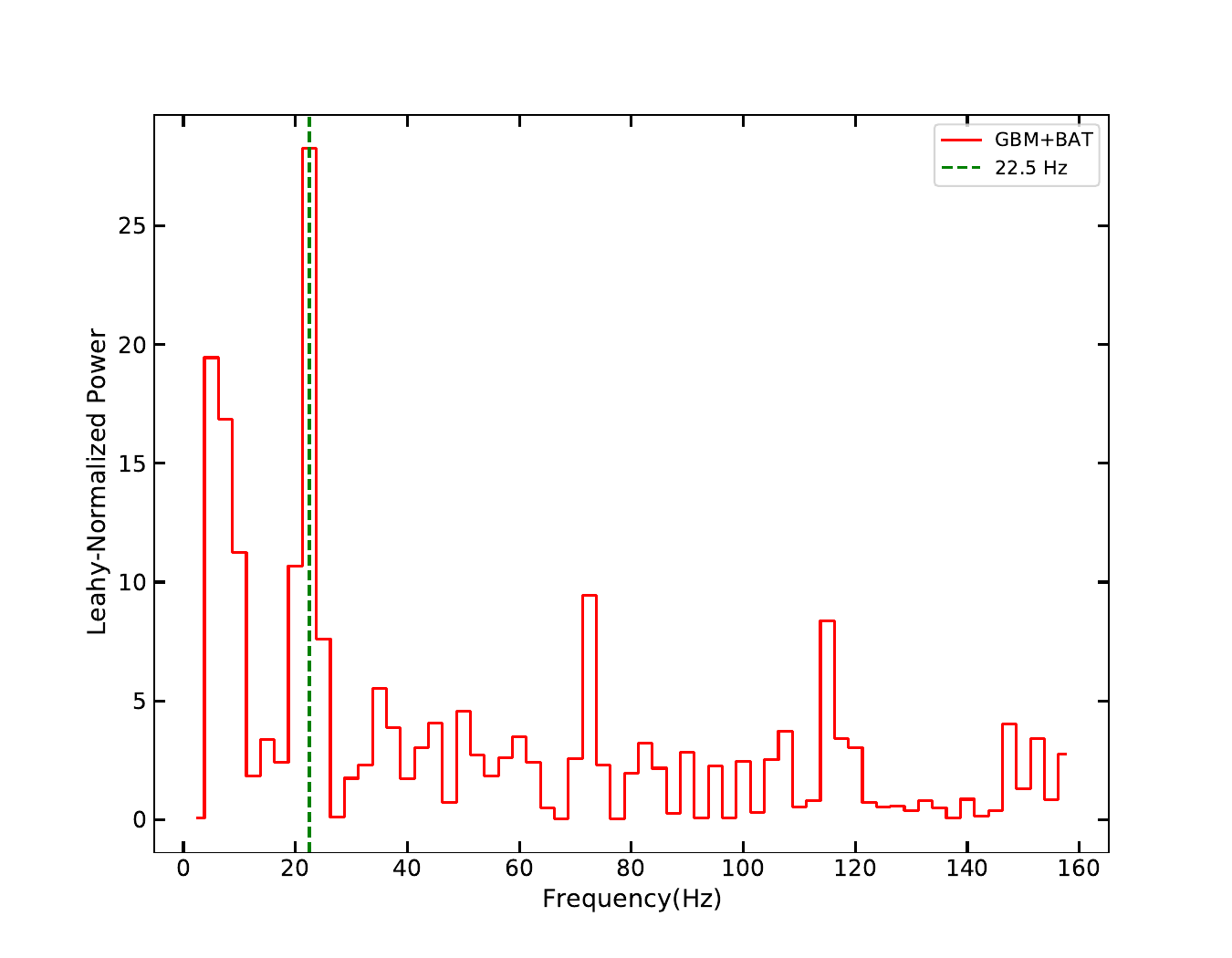}
	\end{minipage}	
	
	\begin{minipage}{0.46\linewidth}
		\centering
        \vskip -0.5cm
        \hspace{-0.3cm}
        \includegraphics[width=8cm,height=6cm]{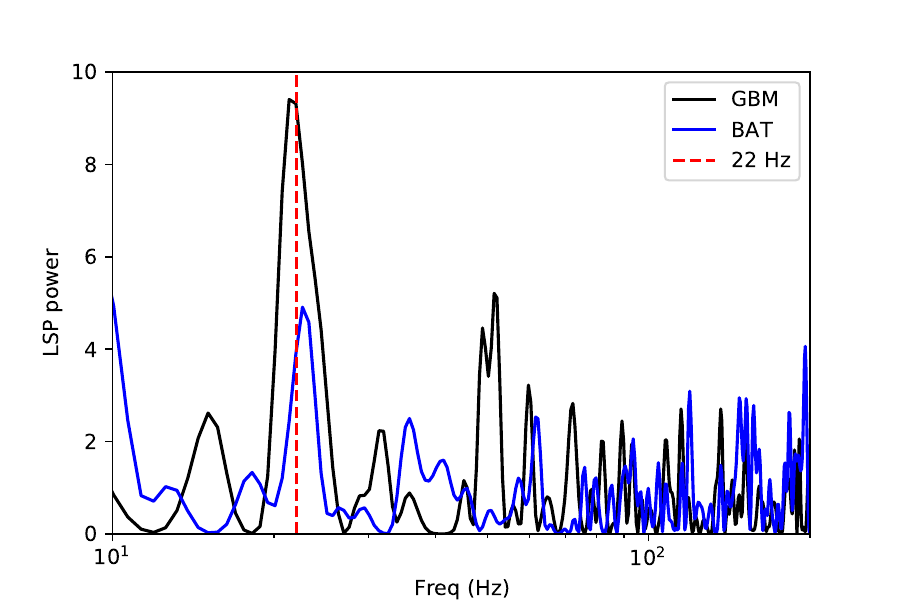}
	\end{minipage}
	\begin{minipage}{0.53\linewidth}
		\centering
        \vskip -0.5cm
        \hspace{-1.3cm}
		\includegraphics[width=\columnwidth]{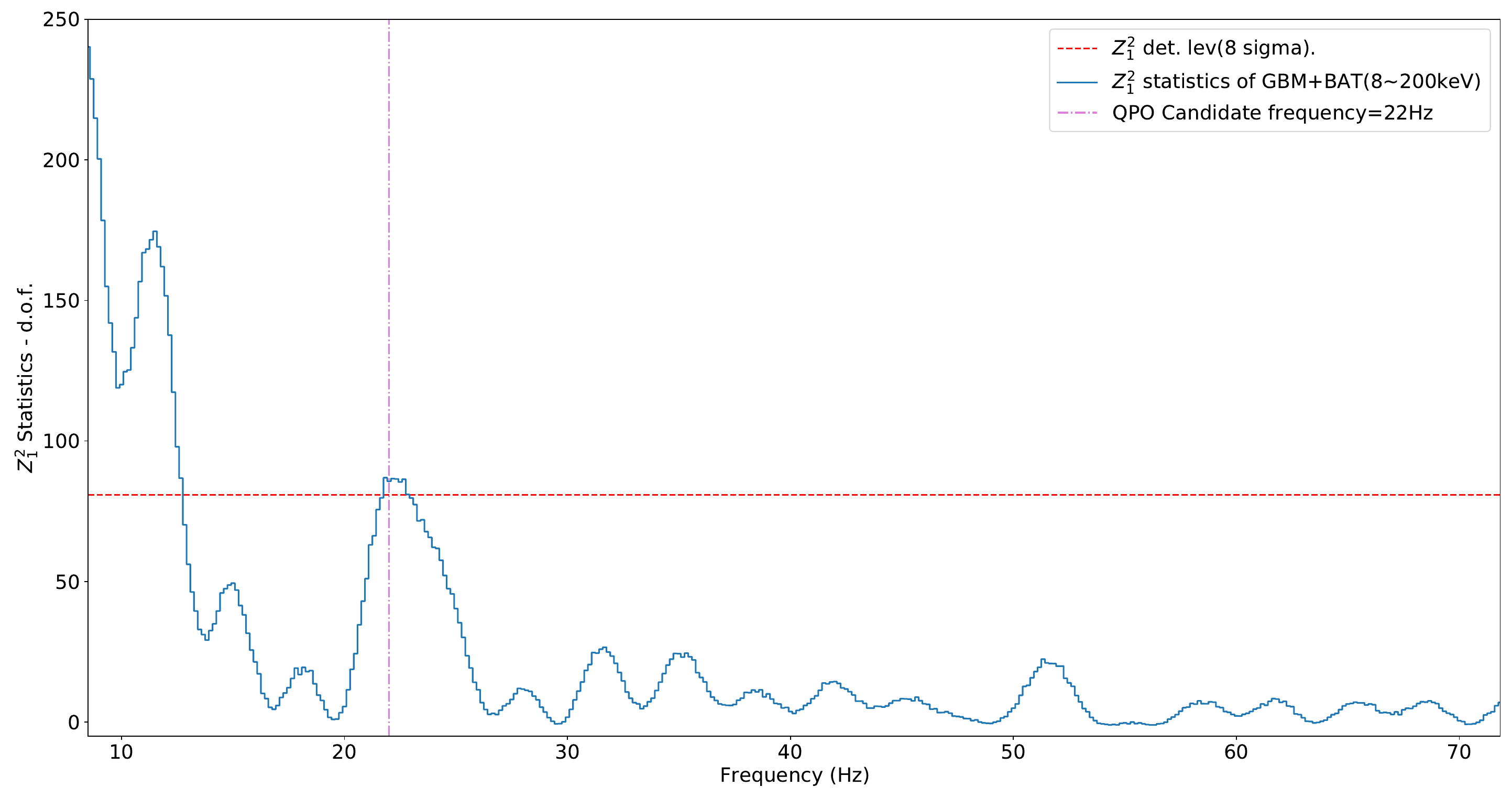}
	\end{minipage}
\caption {Top panels: the lightcurve of GBM and the maximum likelihood fit (top left), and the residual lightcurve minus and the maximum likelihood prediction (top right), the time bin is 8 ms. The significance level is 2.5 $\sigma$. 
Middle panels: The Leahy normalized power of {\it Fermi}/GBM and {\it Swift}/BAT with the lightcurves (middle left) and FRED-model subtracted lightcurves (middle right). The QPO significance is 3.8 $\sigma$ by combining the lightcurves of GBM and BAT.
Bottom left panel: LSP power of GRB\,211211A in 8-200 keV at -0.1 s to 0.3 s. The black and blue solid lines represent LSP power of GBM and BAT, respectively. The red dashed line (22 Hz) represents QPO frequency obtained from WWZ.
Bottom right panel: $Z_{1}^2$ Periodograms for GBM and BAT observations using event by event data, the QPO significance is 8.3 $\sigma$ by combining the data of GBM and BAT. The trials of frequencies searched are considered here.} 
\label{qpo_xb_xb}
\end{figure*}

\begin{figure}
\centering
\includegraphics[width=0.46\textwidth ,height=0.46\textwidth]{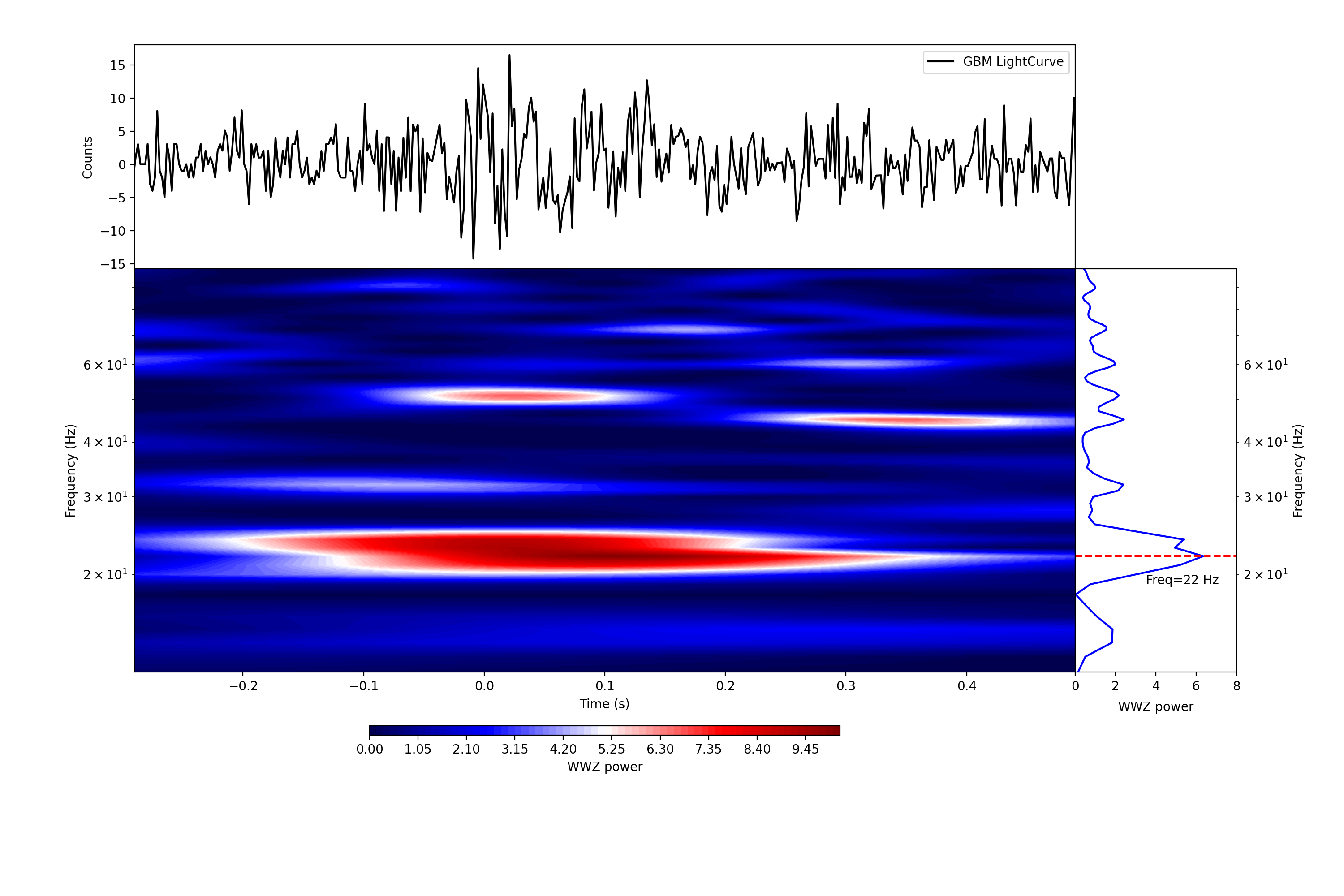}
\includegraphics[width=0.46\textwidth ,height=0.46\textwidth]{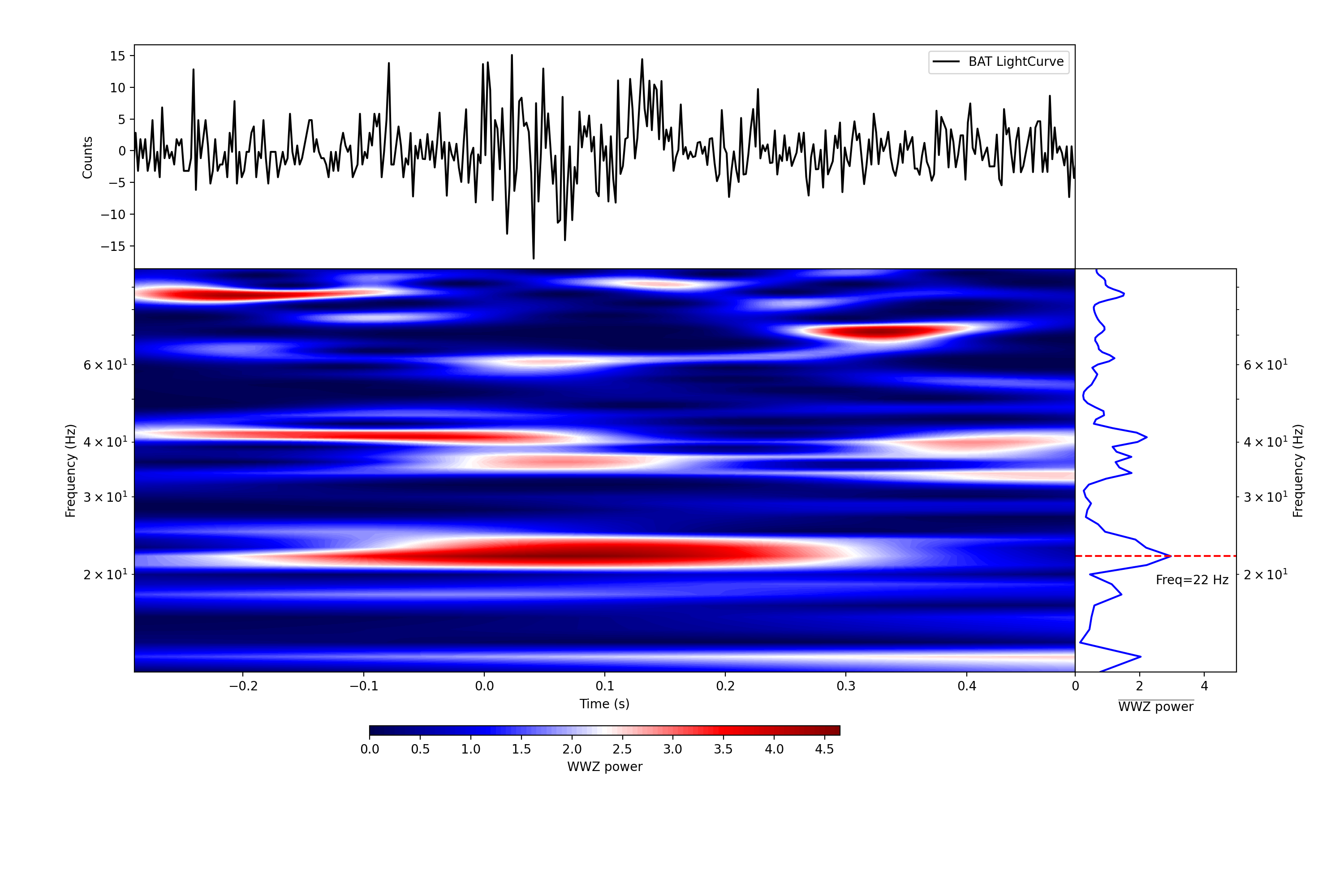}
\caption {FRED-subtracted precursor lightcurve of GBM (upper panel) and BAT (lower panel) in 8-200 keV and the 2D-contour plots of WWZ power. The blue solid lines represent the time-averaged WWZ power respectively. The significance levels for GBM and BAT is 3.0 $\sigma$ and 2.0 $\sigma$, respectively.}
\label{wwz_bat}
\end{figure}

\begin{figure}
	\centering
	\begin{minipage}{1\linewidth}
		\centering
		\includegraphics[width=1\textwidth, height=0.6\textwidth]{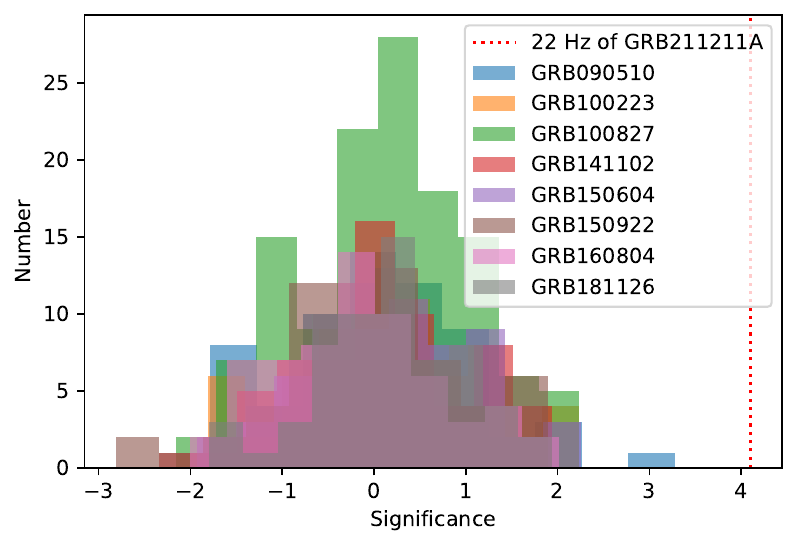}
	\end{minipage}
	\begin{minipage}{1\linewidth}
		\centering
		\includegraphics[width=1\textwidth ,height=0.6\textwidth]{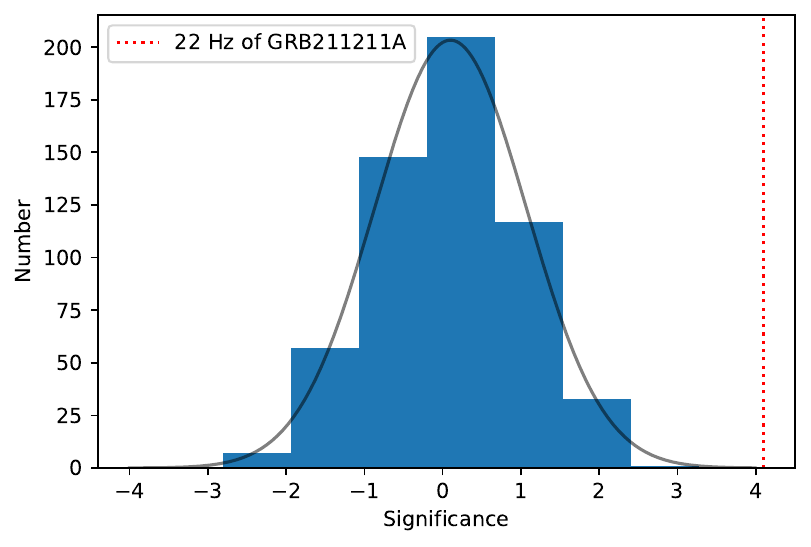}
	\end{minipage}
\caption {Upper panel: the power significance at different frequencies of the other SGRBs precursors with FRED-model subtracted. Lower panel: plot all the histograms in the upper panel into a single histogram, the black line represents the result of fitting a Gaussian function.}
\label{kafang_test}
\end{figure}

\begin{figure*}
\centering
\begin{minipage}[t]{0.46\textwidth}
\centering
\includegraphics[width=\columnwidth]{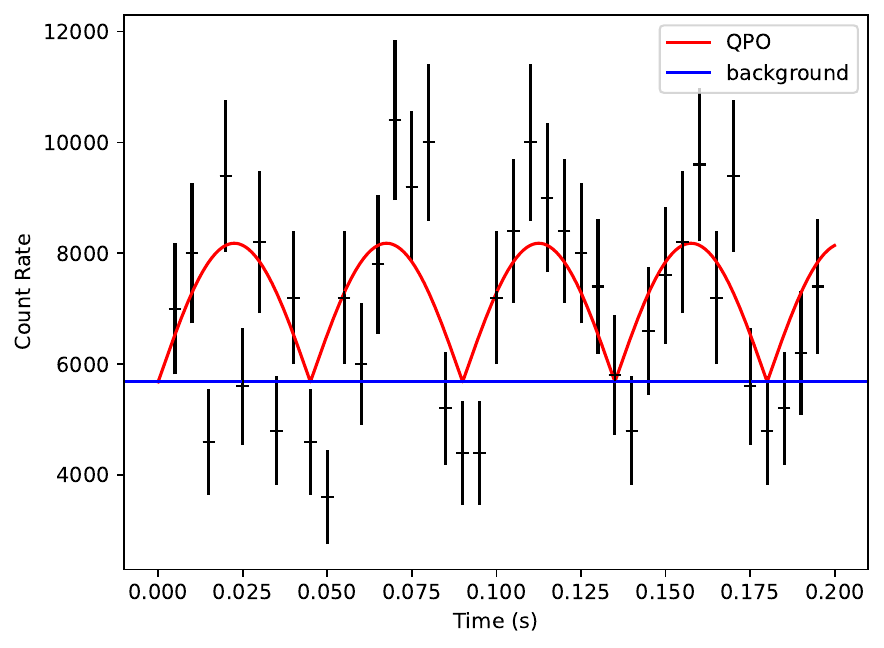}
\end{minipage}
\begin{minipage}[t]{0.46\textwidth}
\centering
\includegraphics[width=\columnwidth]{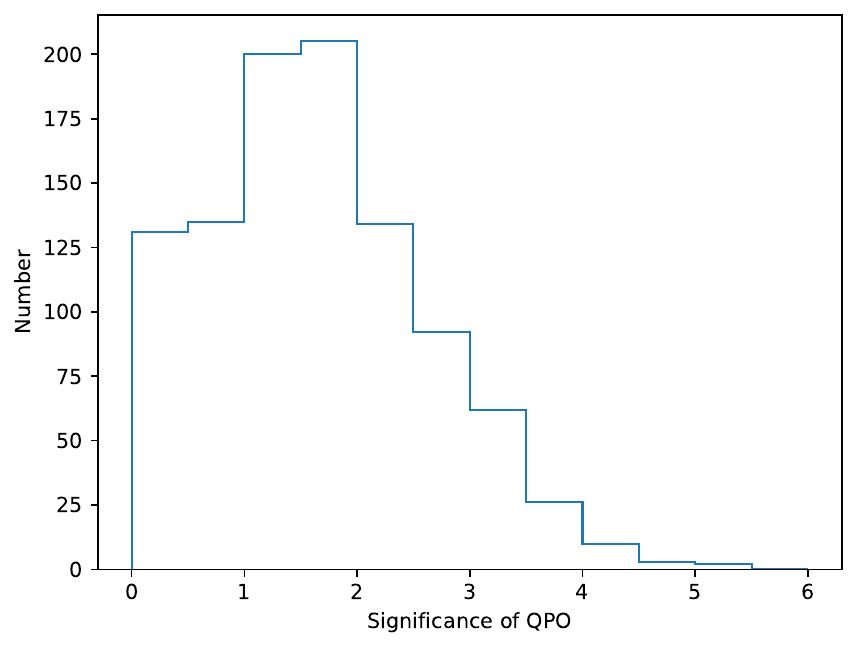}
\end{minipage}
\caption{Left panel: a simulated light curve of the QPO plus the background based on the duration of the precursor in 211211A and the amplitude and frequency of QPO candidate. Right panel: the distribution of QPO significance calculated from the light curves for 1000 simulated observations, there is only two greater than $5\sigma$.}\label{lc_sim}
\end{figure*}

\begin{figure}
\includegraphics[width=\columnwidth]{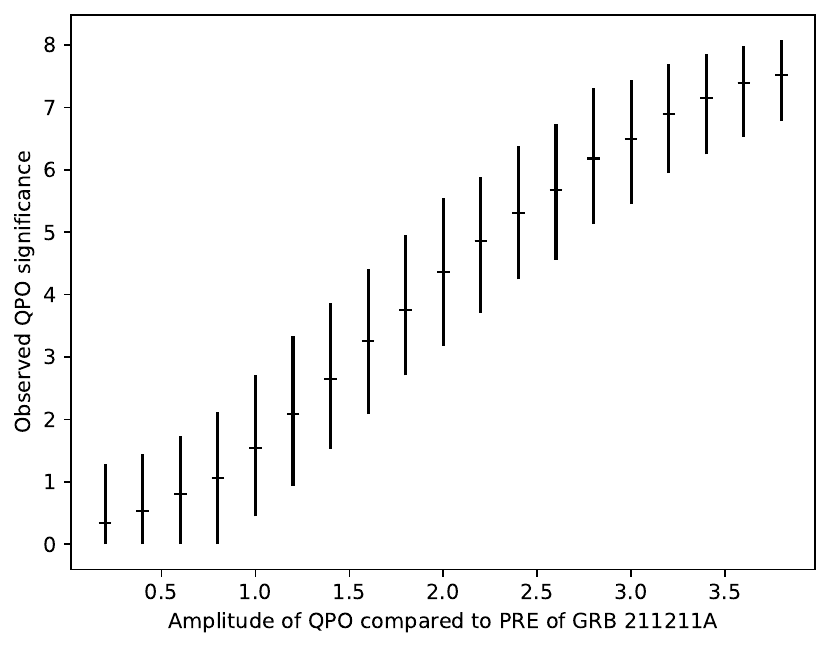}
\caption {The significance of simulated QPO with different amplitudes. The horizontal coordinate represents the QPO amplitude, 1 represents the amplitude of the simulation based on the precursor of GRB 211211A, and the vertical coordinate is the QPO significance, the error bars are $1\sigma$ range.}
\label{fudu_sigma}
\end{figure}

\begin{figure}
	\centering
	\begin{minipage}{1\linewidth}
		\centering
		\includegraphics[width=\columnwidth]{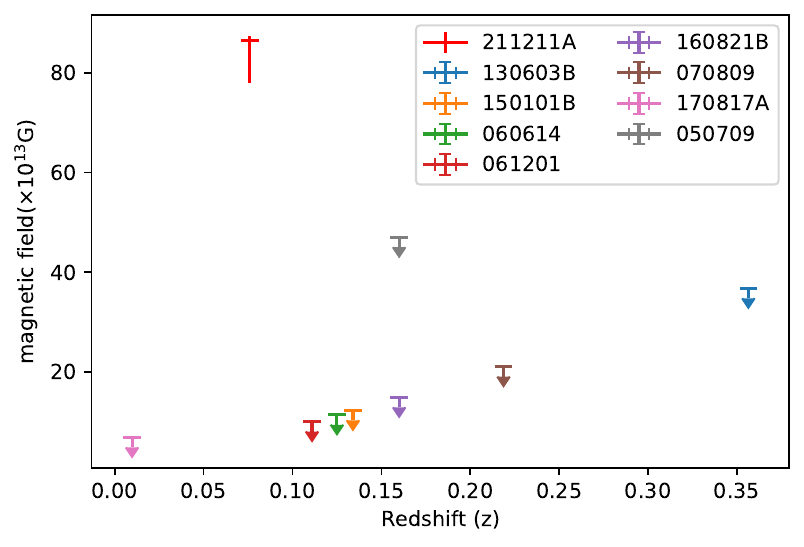}
	\end{minipage}	
\caption {Estimation of neutron star magnetic fields using the pre-merger magnetosphere interaction model \citep{abbott2017gravitational, hansen2001radio} based on the precursors for GRBs with kilonova; the upper limits are calculated based on the sensitivity of the instrument for which no precursors are observed. The magnetic field of GRB 211211A is consistent with the dipole magnetic fields of most known magnetars estimated from their spin-down properties \citep{olausen2014mcgill}.}
\label{magnetat_kilo}
\end{figure}

\begin{table*}
\caption{The spectral fitting with different models of the precursor with {\it Fermi}/GBM and {\it Insight}-HXMT/HE for GRB 211211A.}
\centering
\begin{tabular}{cccccccc}
\hline \hline
Model & $\alpha$ & $E_{\rm cut}$ (keV) & KT (keV) & Flux ($\rm erg/cm^2/s$) & pgstat/dof & BIC \\
\midrule
Comp & $-1.03_{-0.15}^{+0.15}$ & - & $70.74_{-10.81}^{+17.71}$ & $1.76_{-1.04}^{+0.08} \times 10^{-6}$& 376/200 & 397.36 \\
Comp+BB & $-0.81_{-0.10}^{+0.30}$ & $71.88_{-9.38}^{+19.16}$ & $7.73_{-1.37}^{+2.72}$ & $1.80_{-1.18}^{+0.03} \times 10^{-6}$ & 373/198 & 404.95 \\
BB & - & - & $13.32_{-0.49}^{+0.60}$ & $1.51_{-0.11}^{+0.11} \times 10^{-6}$ & 536/201 & 551.84 \\
\hline \hline
\label{fits_models}
\end{tabular}
\end{table*}

\begin{table*}
\centering
\footnotesize
\caption{Spectral fitting with {\it Fermi}/GBM and {\it Insight}-HXMT/HE for GRB 211211A.} \label{fits_table}
\resizebox{\textwidth}{!}{\begin{tabular}{cccccccccc}
\hline \hline
Episode & Model & $E_{\rm cut,1}^*$ (keV) & $\alpha_{1}$ & $\beta$ & $E_{\rm cut,2}^a$ (keV) & $\alpha_{2}$ & $\rm E_{iso}$ ($\rm erg$) & E$^{b}_{\gamma}$ ($\rm erg$) & pgstat/dof \\
\hline
Precursor & Comp & - & - & - & 71 $\pm$ 15 & -1.03$\pm$ 0.16 & $6.97_{-1.29}^{+0.15}$ $\times$ 10$^{48}$ & - & 376/200 \\

Main emission & Band+Comp & 97 $\pm$ 14 & -0.41 $\pm$ 0.14 & -2.04 $\pm$ 0.02 & 1377 $\pm$ 56 & -1.01 $\pm$ 0.04 & 9.02$_{-0.17}^{+0.02}$ $\times$ 10$^{51}$ & 4.06$\times$ 10$^{48}$ & 852/438 \\

Extended emission & Band+Comp & 32 $\pm$ 8 & -0.38 $\pm$ 0.33 & -2.16 $\pm$ 0.04 & 434 $\pm$ 82 & -1.25 $\pm$ 0.16 & $2.60_{-1.49}^{+0.11}$ $\times$ 10$^{51}$ & 1.26$\times$ 10$^{48}$ &  840/807 \\

Full burst & Band+Comp & 63 $\pm$ 10 & -0.56$\pm$ 0.14 & -1.96 $\pm$ 0.02 & 1229 $\pm$ 81 & -1.18 $\pm$ 0.05 & $1.15_{-0.03}^{+0.01}$ $\times$ 10$^{52}$ & - & 1096/388 \\
\hline \hline
\end{tabular}
}

\begin{tablenotes}[normal,flushleft]
\centering
\item{$^a$} $E_{\rm cut}$=$E_{\rm peak}$/(2+$\alpha$). The subscripts 1, 2 represent the parameters of Band and Compton, respectively.
\item{$^b$} Corrected for beaming effect with jet half-opening angle of 0.05 rad.
\item{$^c$} HXMT data are only used for spectral fitting in the precursor and extended emission.

\end{tablenotes} 

\end{table*}

\subsection{QPO candidate}
More interestingly, we find that there are several regularly-spaced pulses superimposed on the FRED trend in the precursor lightcurves (from $T_0$-0.1 s to $T_0$+0.3 s) of \textit{Fermi}/GBM and \textit{Swift}/BAT, which inspired us to search for QPO (see \ref{qpo_xb_xb}).
 We estimate the significance of QPO using Gaussian processes (GP), Z$^2$, fast Fourier transform, weighted wavelet Z-transform (WWZ) and the Lomb-Scargle Periodogram (LSP) methods, respectively. Note that we do not consider the trial numbers of QPO search in other SGRBs with precursors in the estimation of QPO significance, because this GRB 211211A is unique as a long-duration KN-associated merger-origin SGRB.

\subsubsection{Gaussian processes method}
Time-series data can be used to investigate QPO directly in the
time domain using Gaussian Process regression \citep{hubner2022searching}. We model QPOs as a stochastic process on top of a deterministic shape, and perform model selection between QPOs and red noise (see \cite{hubner2022searching} for details). We obtain that $\ln BF_{\rm qpo}$ for GBM data is 2.5. The Bayes factor $BF_{\rm qpo}$ is defined as
\begin{equation}
\begin{split}\label{equ:gs}
BF_{\rm qpo}=\frac{Z(d|k_{\rm qpo+rn},\mu)}{Z(d|k_{\rm rn},\mu)},
\end{split}
\end{equation}
where the numerator (i.e. QPO and red noise) and denominator (i.e. red noise) are the respective evidence in the different models.

We perform this analysis for 1000 simulated light curves produced for each set of parameters and find that the p-value is 0.006, corresponding to 2.5 $\sigma$. 
In this work, we adopt the publicly available code of GP released by \citep{hubner2022searching}.

\subsubsection{The Z$^2$ method}

We also use the GBM data to search for periodic signal with Z-squared ($Z^2$, the harmonic=1) around 25\,Hz in the precursor via $stingray$ \citep{matteo_bachetti_2022_6394742} and find QPOs of $\sim$ 23.49\,Hz with a significance of about 6.6 $\sigma$. The same analyses applied to the BAT data reveal QPOs at 21.87\,Hz with a significance of 4.0 $\sigma$. It is worth noting that the number of trials of frequencies searched is considered throughout this work. The QPO significance is 8.3 $\sigma$ by combining the data of GBM and BAT.
However, we note here that the significance of the $Z^2$ method may be overestimated \citep{hubner2022pitfalls} due to the deviation from the standard $\chi^2-$distribution at each frequency.

\subsubsection{The fast Fourier transform method}
We perform Fast Fourier analysis on the GBM and BAT lightcurves with FRED subtracted, respectively. 
In order to estimate the significance of the 22.5 Hz QPOs, we use Monte Carlo simulations, a widely used tool to evaluate the timing results on a transient event whose profile is known. Specifically, for this precursor, we adopt the FRED function to fit the GBM and BAT lightcurves, and then utilize the Markov Chain Monte Carlo simulations (MCMCs) to generate many sets of FRED parameters of posterior probabilities. With each set of parameters, a FRED profile between -0.1\,s and 0.3\,s can be generated and a simulated lightcurve can be produced by adding the Poisson photon counting noise. Each simulated lightcurve will give a periodogram, and thus one can get the distribution of the periodograms at different Fourier frequencies.

The probability value of the QPOs can be obtained by comparing the Leahy power \citep{van1989fourier} at 22.5\,Hz of the observed light curve to those from the above simulations, which is 0.0016 for GBM and 0.086 for BAT, corresponding to 2.9 $\sigma$ and 1.3 $\sigma$, respectively. The QPO significance is 3.8 $\sigma$ by combining the lightcurves of GBM and BAT. The number of trials of frequencies searched is considered here.

\subsubsection{The weighted wavelet Z-transform method and the Lomb-Scargle Periodogram method}

We first employ one of the most widely used methods, the Weighted Wavelet Z-transform \citep{foster1996wavelets}, to obtain the power spectra of GBM and BAT lightcurves with FRED subtracted. Strong peaks at $22_{-2}^{+3}$ Hz (GBM) and $22_{-1}^{+3}$ Hz (BAT) (with the period cycle of $45_{-4}^{+5}$ ms (GBM) and $45_{-2}^{+5}$ ms (BAT)) (\ref{wwz_bat}) are found in the WWZ power spectra, suggesting the existence of QPOs. The uncertainty of the QPO frequency is the full width at half maximum of the peak. We also use another widely used method, the generalized Lomb-Scargle Periodogram \citep{lomb1976least,zechmeister2009generalised,
scargle1982studies} to obtain the power spectra of GBM and BAT lightcurves, which also show strong peaks at $21.93_{-1.25}^{+2.82}$ Hz (GBM) and $21.62_{-0.63}^{+2.51}$ Hz (BAT) (with period cycle of $45.6_{-5.2}^{+2.8}$ ms (GBM) and $41.5_{-1.4}^{+4.8}$ ms (BAT)) \ref{qpo_xb_xb}. The consistency of the two results also shows the reliability of the analyses. 

To estimate the confidence level of the above QPO signature found with WWZ, we simulate $2 \times 10^{4}$ artificial lightcurves based on GBM and BAT data using the REDFIT method \citep{schulz2002redfit}. The GBM simulation results show that the signal has 3.0 $\sigma$ confidence level, and the BAT simulation results show that the signal has 2.0 $\sigma$ confidence level at $\sim$ 22 Hz. It is worth noting that the REDFIT method needs to satisfy the $\alpha$ of the power law model for the power spectrum is 2. We find $\alpha=2.1\pm 0.1$ (1 $\sigma$) for the precursor of GRB 211211A.

\subsubsection{The power spectrum of other SGRBs precursors with FRED-model subtracted}

To test the validity of using FRED-subtracted precursor lightcurves in the fast Fourier transform, we collect the SGRBs precursors \citep{wang2020stringent} observed in GBM and use the same procedure to analyze them.

First, we find that the SGRBs precursors can be fitted with a single FRED \citep{xiao2022search}. Then we use the same procedure to obtain their power spectra and then obtain the p-values at different frequencies according to a chi-squared distribution with degrees of freedom of 2. The corresponding Gaussian significance is obtained and plotted in the histogram (see top panel of \ref{kafang_test}). We find that 
they follow a Gaussian distribution well, that is, the SGRB precursors do not have significant red noise after subtracting FRED.

\begin{table*}
\centering
\caption{Comparison of the first episodes of bright GRBs 211211A, 130427A, 221009A and 230307A.}
\label{GRBs:precursors}
\resizebox{\textwidth}{!}{
\begin{tabular}{ccccc}
\hline \hline
\thead{GRM name} & \thead{Energy spectrum} & \thead{Spectral lag \\ (between 8-50 keV and 100-300 keV)} & \thead{minimum variability \\ timescale (8-1000 keV)} & \thead{Satisfy the \\ precursor criteria} \\
\hline
GRB 211211A & \thead{CPL, $E_{\rm peak}$=54$\pm$6 keV, \\ $\alpha$=-0.93$\pm$0.28} & 10$\pm$13 ms & 15$\pm$2 ms & Yes \\

GRB 130427A & \thead{SBPL, $E_{\rm break}=291\pm23$ keV, \\ $\lambda_1=-0.334\pm0.025$, \\ $\lambda_2=-3.394\pm0.072$ \citep{Ackermann2014}} & 224$\pm$15 ms & 26$\pm$15 ms & \thead{No (no quiescence \\ period)} \\

GRB 221009A & \thead{CPL, $E_{\rm peak}$=3980$\pm$366 keV, \\ $\alpha$=-1.69$\pm$0.01 \citep{Lesage2023}} & 170$\pm$87 ms & 137$\pm$108 ms & \thead{No (no quiescence \\ period)} \\

GRB 230307A & \thead{Band, $E_{\rm peak}$=170.3$\pm$4.7 keV, \\ $\alpha$=-0.63$\pm$0.04, \\ $\beta$=-2.95$\pm$0.09 \citep{Dichiara2023}} & 12$\pm$2 ms & 5$\pm$1 ms & \thead{No (no quiescence \\ period)} \\
\hline \hline
\end{tabular}
}
\end{table*}

\subsubsection{Significance of simulated light curves}
To understand the significance of this QPO, we perform a simulation. As shown in the top right panel in \ref{qpo_xb_xb}, the maximum amplitude of the QPO obtained by the Gaussian process is about 20 counts per 8 ms (i.e. 2500 counts per second), and we simulate four half-cycles QPO signal with 22 Hz based on the duration ($\sim$ 0.2 s) of the precursor, and we use 5700 per second as the background of GBM. It is worth noting here that due to statistical fluctuations, the light curves may be different for different observations, thus we simulate 1000 light curves and calculate the QPO significance respectively using the fast Fourier transform method method. The left panel of \ref{lc_sim} shows the simulated light curve for a single observation, and the right panel is the distribution of significance of QPO, among of which there are only two cases greater than $5\sigma$. 
Besides, The 68\% confidence interval for this distribution (corresponding to the $1\sigma$ of the Gaussian distribution) is $1.6_{-0.9}^{+1.1}\sigma$, and the QPO significance of the precursor in GRB 211211A observed by GBM is $2.9\sigma$ is within $\sim$ 1.2$\sigma$ of the distribution.
This simulation demonstrates that, even for a real QPO in the observed data, the detection significance of this QPO will not be high in a such short duration precursor and in such low frequency regime as predicted by some magnetar theories (e.g. \citealp{tews2017spectrum, sotani2007torsional}).

We further investigate the significance of QPO with different amplitudes, and repeat the above steps but change the amplitude of QPO. As shown in \ref{fudu_sigma}, the value in the horizontal coordinate represents the QPO amplitude, and value of 1 represents the amplitude of the above simulation based on the precursor of GRB 211211A. The value in vertical coordinate is the QPO significance, and the error bars are $1\sigma$ range. Therefore, we can estimate that the QPO observed with significant above $5\sigma$ is typically more than twice of the amplitude of the precursor of GRB 211211A.
Note that for simplicity, the simulation here does not take into account the possible FRED trend or the evolution of the QPO amplitude with time, which may reduce the significance of QPO.

\subsection{Comparison with other special bursts}
We compared this precursor of GRB 211211A to the first pulse of other bright GRBs (GRB 130427A, GRB 221009A, and GRB 230307A) to assess whether they share common properties. As shown in Table \ref{GRBs:precursors}, only GRB 211211A has a precursor that fully satisfies the precursor criteria. For the spectral properties, the precursor of GRB 211211A is softer than the first pulse of GRB 230307A (see \cite{Dichiara2023} for a detailed analysis of the precursor in GRB 230307A.). For the temporal properties, the spectral lag of the precursor of GRB 211211A is similar to the first pulse of GRB 230307A, while the minimum variability timescale of the precursor of GRB 211211A is slightly larger than that of GRB 230307A. This could be the intrinsic difference between them, but we also note that the minimum variability timescale may be dependent on the signal-to-noise ratio \citep{golkhou2015energy}.
Besides, candidate QPO is only found in the precursor of GRB 211211A.

\section{Discussions}
Considering a binary compact star merger for GRB 211211A, it is difficult to produce strong non-thermal emission in the initial stage of the post-merger phase due to the shielding of ejected material. Indeed, the non-thermal precursors are usually explained by magnetospheric interaction between two neutron stars prior to the merger or the resonant shattering of the crusts of magnetar. Besides, the waiting time between the precursor and main burst ($\sim 1$ s) is consistent with the time interval between the binary neutron star merger (ending of the GW signal) and the GRB in GW 170817 \citep{abbott2017gravitational}. We therefore suggest that this precursor is most likely produced before merger.

Interestingly, if invoking the pre-merger magnetosphere interaction model to explain the luminosity of this precursor, the estimated neutron star magnetic field (using the formula in \citealp{hansen2001radio}) is $\sim 10^{15}$ G (\ref{magnetat_kilo}), which is well consistent with a magnetar. On the contrast, the upper limits for most other GRBs with kilonova (but without a precursor) are lower than the dipole magnetic fields of most known magnetars \citep{olausen2014mcgill}, making this precursor very special. However, this pre-merger magnetosphere interaction model seems difficult to give rise to QPO signal found here.

Despite of many efforts to search for QPOs in GRBs, there are only rare bursts reported before the present work \citep{chirenti2023kilohertz}. By contrast, QPOs of 10 Hz to 1000 Hz have been found in several soft gamma-ray repeaters (SGRs), e.g. Giant Flares (GF) of SGR 1806$-$20 \citep{watts2006detection} and SGR 1900+14 \citep{strohmayer2005discovery}, and an X-ray burst from SGR J1935+2154 and associated with a Fast Radio Burst (FRB 200428) \citep{li2022quasi}. They are usually interpreted as the magnetoelastic or crustal oscillations \citep{link2016torsional,gabler2011magneto,samuelsson2007neutron} of a NS with an extremely strong magnetic field of $\sim 10^{14-15}$ G (i.e. magnetar). From this point of view, the 22 Hz QPO candidate signal reported here may not be a knotty problem, as long as a magnetar is involved in the progenitor system. 

In the magnetar flare scenario, the precursor could be produced by a catastrophic flare accompanying with magnetoelastic or crustal oscillations of the magnetar \citep{zhang2022tidally}. 
For this precursor, the flare might be produced $\sim$ 0.2 s before the coalescence, when the orbital period ($P_{\rm b}$) is $\sim$ 8 ms and the orbital distance is $\sim$ 80 km. Since flares and bursts from magnetars are generally considered to be nearly isotropic, or at least not obviously beamed, it is interesting to note that the precursor's total energy output $E_{\rm iso }\sim 7.7\times 10^{48}$ erg is comparable to the jet-corrected energies $E_{\gamma}$ for ME and EE, which are $6.6\times10^{48} $ erg and $2.8\times10^{48} $ erg, respectively. Therefore this precursor of GRB 211211A is energetically very important for this event.

The involvement of magnetar in the merger system may also be supported by the prolonged burst time of GRB 211211A. Numerical relativity simulations show that for the binary merger of highly magnetized NSs, the central engine of a GRB would be modeled by a magnetized accretion torus with saturated magnetic field strength \citep{kiuchi2014high}. With the magnetic barrier effect, radial angular momentum transfer may significantly prolong the lifetime of the accretion process \citep{proga2006late, liu2012radial}, consistent with the long duration of GRB 211211A. 

However, there appears to be a challenge between the relatively short lifetime of a magnetar and the much longer spiraling time ($>$ 86 Myr) of a binary NS system due to the GW radiation \citep{Tauris2017}. If the former is much smaller than the latter, it implies that the previous magnetar will be no longer highly magnetized in the late inspiral phase. However, estimating the lifespan of a magnetar and the spiraling time of a binary NS system involves many uncertainties. Magnetar age is usually $<\sim 10^{4}$ years estimated from their observed period and period derivative as well as the decay of the magnetic field \citep{kaspi2017magnetars}. If only consider the decay of the magnetic field, it is usually $<\sim 10^{7}$ years \citep{kaspi2017magnetars,Beniamini19}. One potentially relevant object is SGR 0418+5729 \citep{Rea2013}, which is apparently much older ($\sim 3.6\times 10^{7}$ years \citep{Beniamini19}), but it may still be not old enough to explain GRB 211211A.
However, there is no consensus on the timescale of magnetic field decay. Additionally, the spiraling time is heavily influenced by the initial orbit of the inspiral during its formation and the environment surrounding the binary system. Another possibility is that the weak magnetic field of a NS can be amplified up to a strength of $\sim 10^{15}$ G during the late spiraling phase, which means that an old NS may be recycled into a magnetar in this process \citep{osso2013}. 

In addition, as the binary period deceases, crust oscillations could become stronger and stronger. From the top right panel in \ref{qpo_xb_xb}, the QPO amplitude seems to firstly increase slightly and then decreases slowly (similar to the QPO in another magnetar burst \citep{li2022quasi}); this is not fully consistent with the aforementioned expectation if the observed QPO amplitude is positively correlated with the strength of the crust oscillations. On the other hand, it is unclear how the QPO amplitude in X-ray/Gamma-ray flux is related to the strength of the crust oscillations, which also requires more theoretical calculations.

\section{Summary}
In this work, we report a peculiar precursor in the GRB 211211A, whose temporal and spectral characteristics are different from that of the main emission and extended emission of GRB 211211A. We find that this precursor is a rare case with high-confident nonthermal spectrum in a GRB originating from a binary merger. 

In particular, we find that there is a QPO candidate in the precursor. We utilize a series of temporal and frequency domain QPO search methods, including GP, Z$^2$, fast Fourier transform and WWZ methods, to calculate the significance, and obtain 2.5 $\sigma$, 6.6 $\sigma$, 2.9 $\sigma$ and 3.0 $\sigma$ observed by GBM, respectively. Considering all these results, we summarize that the significance of the QPO candidate should be $\sim 3\sigma$. We note that the QPO significance ($\sim 7\sigma$) found in GRB 200415A using the Z$^2$ method \citep{castro2021very} is close to that (6.6 $\sigma$) of this precursor in GRB 211211A. In addition, we searched QPO in the precursors of the other short GRBs by the same methods but do not find any significant signal, that is, the significance of the QPO candidate in the precursor of GRB 211211A is the highest among all precursors of short GRBs. Moreover, our simulations illustrate that it is difficult to obtain highly significant low-frequency QPO in such a short precursor, lending additional support to the QPO signal found in GRB 211211A.

 We suggest that this peculiar precursor could be explained either by the pre-merger magnetosphere interaction model or the magnetar crustal oscillations model. For the former, the magnetic field of the neutron star is estimated to be $\sim 10^{15}$ G, which is well consistent with a magnetar. For the latter, a magnetar catastrophic flare (or called superflare) can also well explain the characteristics of the precursor, especially the QPO feature \citep{zhang2022tidally}. Thus, we conclude that both models point towards to the magnetar involvement in the merger. Interestingly, the strong magnetic field of the magnetar can naturally lead to the prolonged duration of GRB 211211A, which is otherwise difficult to explain. Meanwhile, more theoretical discussions have been proposed to interpret the observation features of this peculiar precursor \citep{zhang2022tidally,Suvorov2022via,sullivan2024gamma}.

Finally, gravitational waveforms of magnetized and non-magnetized NS binaries could be well distinguished as long as the NS magnetic field is strong enough \citep{giacomazzo2009can}. Future multi-messenger observations have the potential to confirm the presence of highly magnetized NS as progenitors of bursts similar to GRB 211211A, and to determine if their precursors are produced by magnetar flares. Furthermore, these observations could provide valuable insights into the physical mechanism underlying the precursors and their intriguing behaviors, including possible QPOs.

\section{Acknowledgments}
This work made use of the data from the {\it Insight}-HXMT, {\it Fermi} and {\it Swift}. This work is supported by the National Key R\&D Program of China (2021YFA0718500), the National Natural Science Foundation of China (NO. 12303043, 12273042, Projects: 12061131007, 12021003), the Strategic Priority Research Program (Grant No. 
XDA30050000, 
XDB0550300,  
XDA15052700, XDB23040400) of the Chinese Academy of Sciences,
and the International Partnership Program of Chinese Academy of Sciences (Grant No.113111KYSB20190020).
D.X. acknowledges the support by the Strategic Priority Research Program “Multi-wavelength Gravitational Wave Universe” of the CAS (No. XDB23000000) and the science research grants from the China Manned Space Project with NO. CMS-CSST-2021-A13, CMS-CSST-2021-B11. 
We thank Xing Gao for kind help of the Nanshan/NEXT observations. We thank Bing Zhang and Daniela Huppenkothen for helpful discussions. We acknowledge the support of the staff of the Xinglong 2.16m telescope. This work was also partially supported by the Open Project Program of the Key Laboratory of Optical Astronomy, National Astronomical Observatories, Chinese Academy of Sciences.

\bibliography{main}

\end{CJK}
\end{document}